\documentclass[12pt,preprint]{aastex}

\newcommand\iso[2]{$^{\rm #1}$#2}

\shorttitle{$\omega$ Centauri Abundances}
\shortauthors{Johnson et al.}

\begin{document}

\title{A Large Sample Study of Red Giants in the Globular Cluster Omega
Centauri (NGC 5139)}

\author{Christian I. Johnson\altaffilmark{1}, 
Catherine A. Pilachowski\altaffilmark{1}, R. Michael Rich\altaffilmark{2}, and
Jon P. Fulbright\altaffilmark{3,4}
}

\altaffiltext{1}{Department of Astronomy, Indiana University,
Swain West 319, 727 East Third Street, Bloomington, IN 47405--7105, USA;
cijohnson@astro.indiana.edu; catyp@astro.indiana.edu}

\altaffiltext{2}{Department of Physics and Astronomy, UCLA, 430 Portola Plaza,
Box 951547, Los Angeles, CA 90095-1547, USA; rmr@astro.ucla.edu}

\altaffiltext{3}{Department of Physics and Astronomy, Johns Hopkins University,
Baltimore, MD 21218, USA; jfulb@skysrv.pha.jhu.edu}

\altaffiltext{4}{Visiting astronomer, Cerro Tololo Inter--American 
Observatory, National Optical Astronomy Observatory, which are operated by the 
Association of Universities for Research in Astronomy, under contract with the 
National Science Foundation.}

\begin{abstract}

We present abundances of several light, $\alpha$, Fe--peak, and 
neutron--capture elements for 66 red giant branch (RGB) stars in the Galactic 
globular cluster Omega Centauri ($\omega$ Cen).  Our observations lie in the 
range 12.0$<$V$<$13.5 and focus on the intermediate and metal--rich RGBs.  
Abundances were determined using equivalent width measurements and spectrum
synthesis analyses of moderate resolution (R$\approx$18,000) spectra obtained
with the Blanco 4m telescope and Hydra multifiber spectrograph.  Combining 
these data with previous work, we find that there are at least four peaks in 
the metallicity distribution function at [Fe/H]=--1.75, --1.45, --1.05, and 
--0.75, which correspond to about 55$\%$, 30$\%$, 10$\%$, and 5$\%$ of our 
sample, respectively.  Additionally, the most metal--rich stars are the most 
centrally located.  Na and Al are correlated despite exhibiting star--to--star 
dispersions of more than a factor of 10, but the distribution of those 
elements appears to be metallicity dependent and are divided at 
[Fe/H]$\approx$--1.2.  About 40--50$\%$ of stars with [Fe/H]$<$--1.2 have Na 
and Al abundances consistent with production solely in Type II supernovae and 
match observations of disk and halo stars at comparable metallicity.  The 
remaining metal--poor stars are enhanced in Na and Al compared to their disk 
and halo counterparts and are mostly consistent with predicted yields from 
$>$5 M$_{\sun}$ asymptotic giant branch (AGB) stars.  At [Fe/H]$>$--1.2, more 
than 75$\%$ of the stars are Na/Al enhanced and may have formed almost 
exclusively from AGB ejecta.  Most of these stars are enhanced in Na by at 
least 0.2 dex for a given Al abundance than would be expected based on 
``normal" globular cluster values.  All stars in our sample are $\alpha$--rich
with $\langle$[Ca/Fe]$\rangle$=+0.36 ($\sigma$=0.09) and 
$\langle$[Ti/Fe]$\rangle$=+0.23 ($\sigma$=0.14).  The Fe--peak elements give
solar--scaled abundances and similarly small dispersions with 
$\langle$[Sc/Fe]$\rangle$=+0.09 ($\sigma$=0.15) and
$\langle$[Ni/Fe]$\rangle$=--0.04 ($\sigma$=0.09).  Europium does not vary
extensively as a function of metallicity and has 
$\langle$[Eu/Fe]$\rangle$=+0.19 ($\sigma$=0.23).  However, [La/Fe] varies from 
about --0.4 to +2 and stars with [Fe/H]$\ga$--1.5 have [La/Eu] values 
indicating domination by the s--process.  A quarter of our sample have 
[La/Eu]$\geq$+1 and may be the result of mass transfer in a binary system.  We 
conclude that the metal--rich population must be at least 1--2 Gyr younger 
than the metal--poor stars, owing to the long timescales needed for strong 
s--process enrichment and the development of a large contingent of mass 
transfer binaries.

\end{abstract}


\section{INTRODUCTION}

Among all of the known Galactic globular clusters, Omega Centauri ($\omega$ 
Cen) is unique in the extent of its chemical enrichment.  The cluster exhibits 
huge star--to--star abundance variations that are not limited solely to the 
light elements, as is the case for most ``normal" globular clusters.  Instead, 
$\omega$ Cen stars have [X/Fe]\footnote{We make use of the standard 
spectroscopic notations where 
[A/B]$\equiv$log(N$_{\rm A}$/N$_{\rm B}$)$_{\rm star}$--
log(N$_{\rm A}$/N$_{\rm B}$)$_{\sun}$ and 
log $\epsilon$(A)$\equiv$log(N$_{\rm A}$/N$_{\rm H}$)+12.0 for elements A
and B.} dispersions of 0.5 to more than 1.0 dex for many elements and span a 
metallicity range from [Fe/H]$\approx$--2.2 to nearly --0.5  (e.g., Norris \& 
Da Costa 1995; Suntzeff \& Kraft 1996; Smith et al. 2000; Johnson et al. 2008).
Additionally, $\omega$ Cen's red giant branch (RGB) and subgiant branch (SGB)
show 4--5 discrete populations in concert with multiple main sequences (Lee et 
al. 1999; Hilker \& Richtler 2000; Pancino et al. 2000; van Leeuwen et al. 
2000; Ferraro et al. 2004; Rey et al. 2004; Stanford 2004, 2006; Sollima et 
al. 2005a; Villanova et al. 2007).  These data, along with the apparent 
age dispersion at the main sequence turnoff, suggest $\omega$ Cen underwent 
extensive self--enrichment and star formation over $>$1 Gyr.

While $\omega$ Cen has an estimated mass of $\sim$2--7$\times$10$^{\rm 6}$ 
M$_{\sun}$ (Richer et al. 1991; Meylan et al. 1995; van de Ven et al. 2006), 
it does not appear to have a particularly deep potential well compared to 
other lower mass clusters (Gnedin et al. 2002).  Combined with the cluster's
retrograde orbit and short disk crossing time ($\sim$1--2$\times$10$^{\rm 8}$
yrs; Dinescu et al. 1999), it seems unlikely that star formation could have 
occurred over several Gyrs in the cluster's current configuration.  A proposed 
scenario is that $\omega$ Cen was once the nucleus of a dwarf spheroidal
galaxy that was accreted and stripped apart via gravitational interaction with
the Milky Way (e.g., Bekki \& Norris 2006).  If this is true, then the cluster 
was probably much more massive in the past.

Until recently, $\omega$ Cen was the only known globular cluster with multiple
populations, but new observations (e.g., Piotto 2008) have indicated several
of the more massive clusters in the Galaxy host at least two SGBs and/or main
sequences despite being monometallic.  These anomalous sequences are often 
interpreted as having large He enhancements ranging from Y$\sim$0.30--0.38, 
compared to the canonical He abundance of Y$\sim$0.25.  This assumption 
applies to the blue main sequence in $\omega$ Cen as well, which is roughly 
0.3 dex more metal--rich than the red main sequence and requires Y$\sim$0.38 to match the observations in this paradigm (Bedin et al. 2004; Norris 2004;
Piotto et al. 2005).  However, the important caveat remains that while the
metallicity difference is \emph{measured}, the He difference is only
\emph{inferred}.  The source of these potential He enhancements remains
unknown, but the most likely candidates include: 3--8 M$_{\sun}$ asymptotic 
giant branch (AGB) stars, super--AGB stars ($\sim$8--10 M$_{\sun}$), massive 
rotating stars, and population III stars (e.g., see Renzini 2008 for a review 
of this topic).  Each of these scenarios poses a unique set of obstacles,
but the basic problem is the difficulty in producing a discrete 
\emph{population} of He--enriched stars while satisfying other chemical, age, 
and IMF constraints.

Globular cluster stars appear to have a more complex chemical history than
their halo counterparts of similar metallicities, particularly with
respect to the light elements oxygen through aluminum.  In moderately
metal--poor halo stars (--2.0$\la$[Fe/H]$\la$--1.0), these elements closely
mimic the trends predicted for stars forming primarily out of gas polluted by
core collapse supernovae (SNe; e.g., Timmes et al. 1995; Samland 1998; Nomoto
et al. 2006).  That is, the $\alpha$ elements remain enhanced at 
[$\alpha$/Fe]$\sim$+0.40, but Na and Al, due to their secondary (i.e., 
metal--dependent) production, slowly increase relative to Fe with increasing 
metallicity.  This is contrasted with the ubiquitous trends observed in 
globular clusters, which have stars with similar abundance patterns (the 
so--called ``primordial" population) \emph{and} stars showing signs of varying 
degrees of high temperature proton--capture processing (the ``intermediate" 
and ``extreme" populations; e.g., Kraft 1994; Gratton et al. 2004; Carretta et 
al. 2008).  These tell--tale signs of additional processing are evidenced by 
the pervasive O--Na and Mg--Al anticorrelations along with the Na--Al 
correlation observed in all well--studied clusters to date, and are the result
of processing in the ON, NeNa, and MgAl cycles (e.g., Gratton et al. 2004).  
Since these trends are observed in main sequence and turnoff stars (Cannon et 
al. 1998; Gratton et al. 2001; Cohen et al. 2002; Briley et al. 2004a; 2004b; 
Boesgaard et al. 2005) as well as RGB stars, it seems likely that the chemical 
patterns were already imprinted in the gas from which the current generation 
of stars formed.  The source of these abundance patterns is unknown, but 
intermediate mass AGB stars, which undergo hot bottom burning (HBB) at 
temperatures exceeding 80--100$\times$10$^{\rm 6}$ K and experience 
third dredgeup, are a popular choice because they do not alter [Fe/H],
have low velocity ejecta, and produce large quantities of He, thus possibly
alleviating some of the He enhancement issues mentioned above.  While AGB 
stars are a qualitatively attractive solution, many problems arise in 
quantitative analyses and the ejecta yields are highly model dependent (e.g., 
Denissenkov \& Herwig 2003; Fenner et al. 2004; Choi \& Yi 2008; Ventura \& 
D'Antona 2008).  Other potential polluters include fast rotating massive stars 
(Decressin et al. 2007) and previous generations of slightly more massive RGB 
stars (Denissenkov \& Weiss 2004); \emph{In situ} deep mixing may also still 
play a role in highly evolved RGB stars.

In terms of chemical properties, $\omega$ Cen behaves similarly to Galactic
globular cluster populations (aside from the large metallicity spread) in that 
the various light element relations and $\alpha$ enhancements are present in 
nearly all subpopulations analyzed so far (e.g., Norris \& Da Costa 1995; Smith
et al. 2000), but the cluster hosts stars of considerable Na/Al enrichment and 
O depletion.  Unlike the field and disk populations that exhibit lower 
[$\alpha$/Fe] ratios at [Fe/H]$>$--1, presumably due to the contributions of 
Type Ia SNe, the overwhelming majority of $\omega$ Cen stars at the same
metallicity are $\alpha$ enhanced.  This suggests that Type Ia SNe have played 
only a minor role in the cluster's chemical enrichment for all but perhaps the 
most metal--rich stars (Pancino et al. 2002; but see also Cunha et al. 2002).  
If $\omega$ Cen is the remnant of a former dwarf spheroidal galaxy then it has 
evolved much differently than present day dwarf galaxies because they do not 
show extreme light element enhancements/depletions and often exhibit subsolar 
[$\alpha$/Fe] abundances (e.g., see review by Geisler et al. 2007).  However, 
$\omega$ Cen does share the stronger s--process component seen in many dwarf 
spheroidal stars, except that the cluster stars more metal--rich than 
[Fe/H]$\sim$--1.5 show s--/r--process ratios indicating complete s--process 
dominance whereas the dwarf galaxies experienced much weaker s--process 
enrichment.  This is in direct contradiction to globular clusters, which are 
r--process rich.  Lower mass AGB stars ($\sim$1--4 M$_{\sun}$), which are 
thought to produce most of the s--process elements, have therefore had a much 
more significant effect on $\omega$ Cen's chemical evolution than is seen in 
dwarf spheroidals and globular clusters.  

In this paper we present spectroscopic analyses of numerous light, $\alpha$, 
Fe--peak, and neutron--capture elements for 66 stars spanning $\omega$ Cen's 
full metallicity range, with an emphasis on the lesser studied intermediate
and metal--rich populations.  We combine our results with those from the 
literature and compare $\omega$ Cen to the Galactic thin and thick disk, 
halo, bulge, other globular clusters, and nearby dwarf spheroidals in an
attempt to disentangle the evolution of these very different populations and
perhaps isolate chemical signatures that are unique to each subpopulation
in $\omega$ Cen.

\section{OBSERVATIONS AND REDUCTIONS}

All observations were taken at the Cerro Tololo Inter-American Observatory on
2006 May 26 and 2006 May 27 using the Blanco 4m telescope and Hydra multifiber 
spectrograph.  In each configuration we used the ``large" 300$\micron$ 
(2$\arcsec$) fibers and obtained spectra with two different bench spectrograph 
setups.  The first setup was centered near 6670~\AA\ and spanned approximately 
6530--6800~\AA\ while the second setup was centered on 6125~\AA\ and ranged 
from 6000--6250~\AA.  Both spectrograph setups employed the 100$\micron$ slit 
mask along with the 400 mm Bench Schmidt Camera and 316 line mm$^{\rm -1}$ 
echelle grating to achieve a resolving power of 
R($\lambda$/$\Delta$$\lambda$)$\approx$18,000 (0.35~\AA\ FWHM).  

Target stars, coordinates, photometry, and membership probabilities were taken
from the proper motion study by van Leeuwen et al. (2000).  The targets were
chosen to be on the upper third of the giant branch and all have V$<$14.0, but
priority was given to those with larger B--V indices (i.e., more metal--rich)
in the Hydra assignment code.  Stars with membership probabilities $<$70$\%$
were excluded from the fiber assignment process.  While we did not measure 
radial velocities for the target stars, cluster members were easily discerned
from the field star population because of $\omega$ Cen's comparatively large 
radial velocity ($\langle$V$_{\rm R}$$\rangle$$\sim$232 km s$^{\rm -1}$; Reijns
et al. 2006).

\begin{figure}
\epsscale{1.00}
\plotone{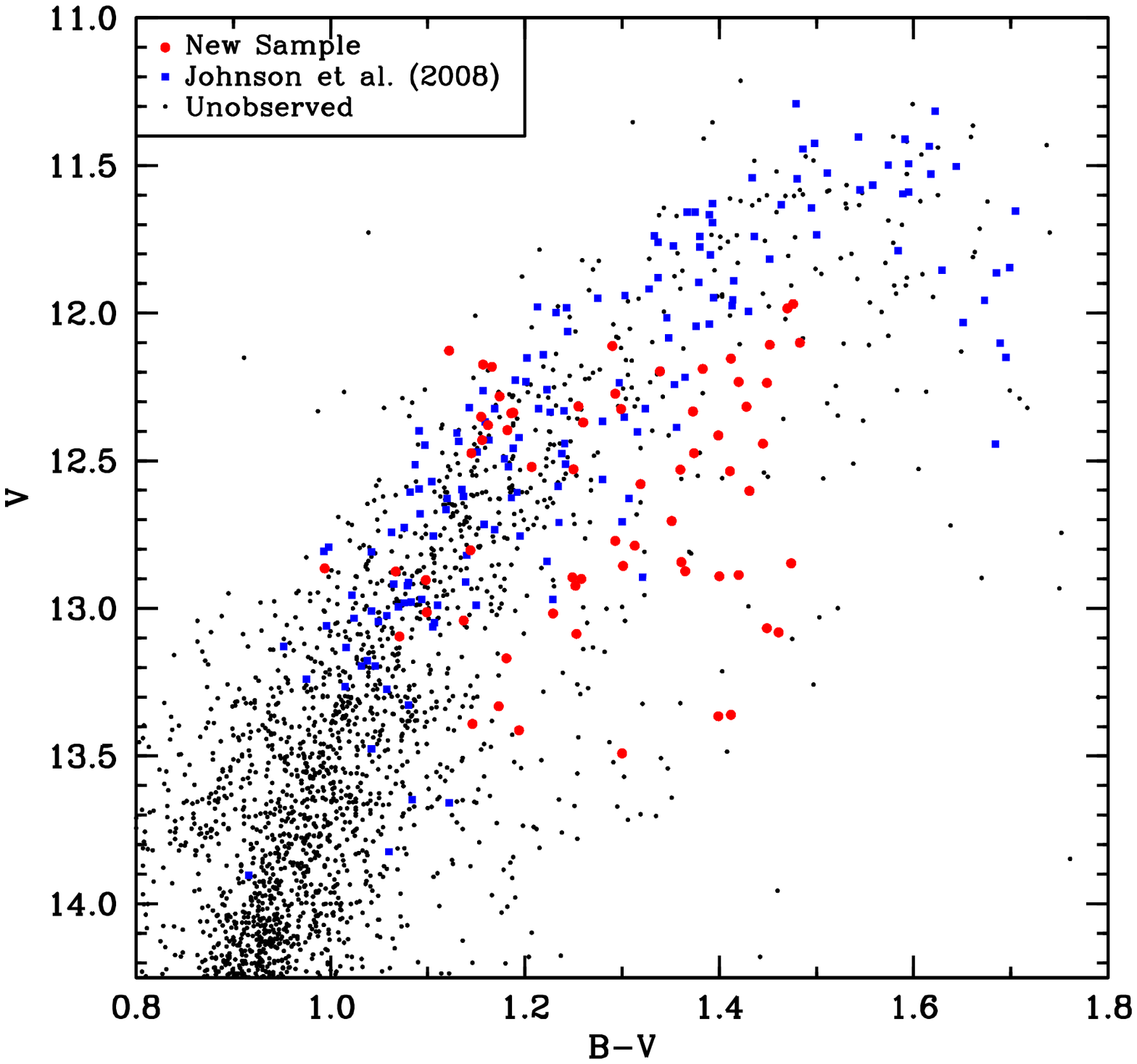}
\caption{Color--magnitude diagram of $\omega$ Cen's RGB.  The filled red 
circles represent the stars observed for this study and the filled blue squares
show the stars observed for Johnson et al. (2008).  There are 22 stars which
overlap with Johnson et al. and those are also indicated by filled red circles.
The complete sample, including stars not observed here, are from van Leeuwen 
et al. (2000).}
\label{f1}
\end{figure}

We obtained 3, 1800 second exposures for each spectrograph setup with 92 
fibers placed on targets.  The co--added signal--to--noise (S/N) ratios of the 
spectra ranged from $\sim$25--200, but we only analyzed stars for which the 
S/N was $\ga$50.  The final sample includes 66 stars and are shown in 
Figure \ref{f1} along with the data from Johnson et al. (2008) and the full 
sample of van Leeuwen et al. (2000).

\begin{figure}
\epsscale{1.00}
\plotone{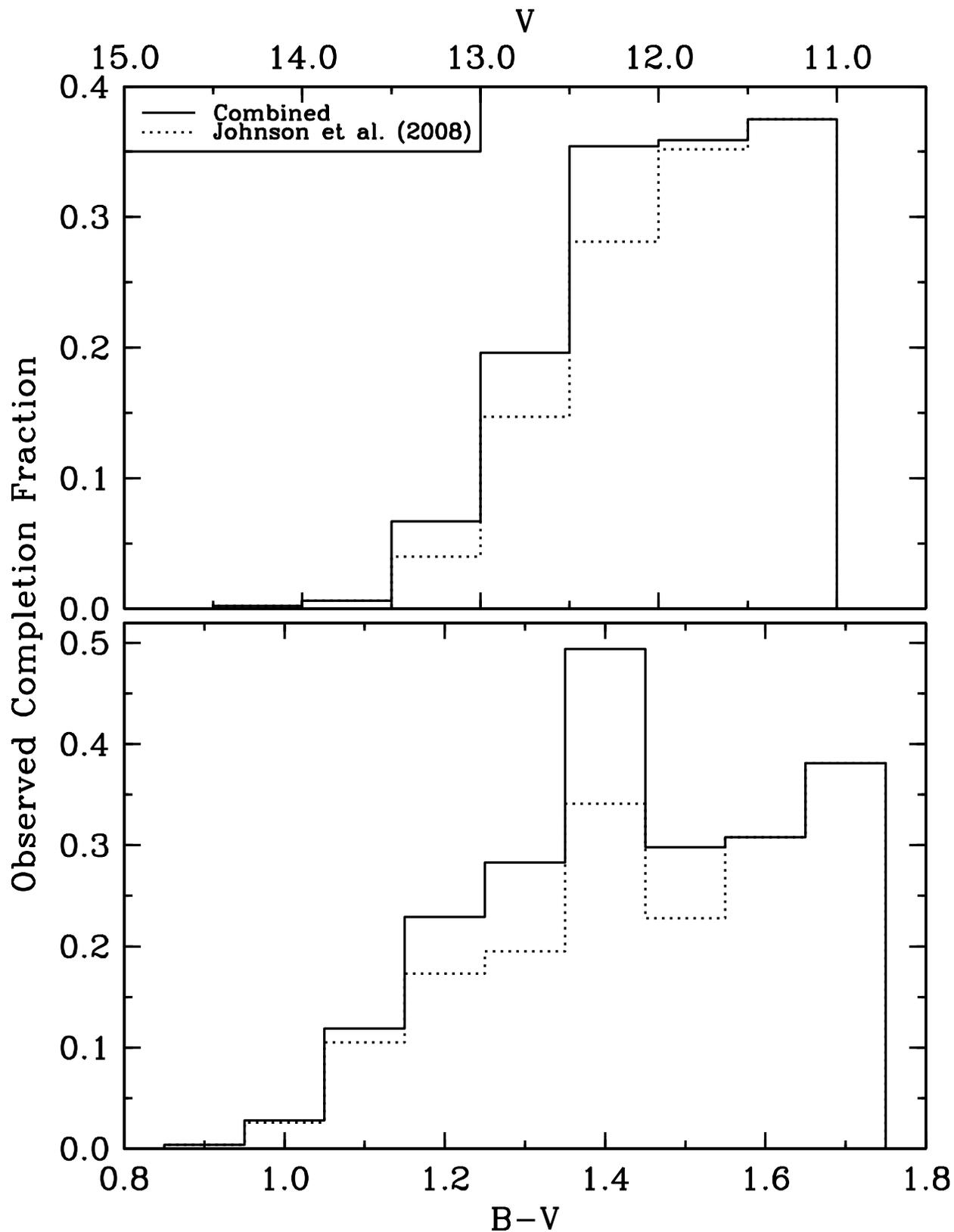}
\caption{Histogram showing the observed completion fraction of this study 
combined with the data of Johnson et al. (2008).  The top panel shows the 
completion fraction as a function of V magnitude and the bottom panel shows
the completion fraction as a function of B--V color.}
\label{f2}
\end{figure}

Since $\omega$ Cen exhibits such a large range in metallicity and the various
giant branches contain stars in different ratios, selection effects may be 
more prominent than for typical globular clusters.  In Figure \ref{f2} we show 
the observed completion fractions of our current data combined with Johnson 
et al. (2008) as a function of both V magnitude and B--V color.  While there was
little increase in the completion fraction for stars with 11.0$<$V$<$12.0, 
those with 12.0$<$V$<$13.5 increased $\sim$5-10$\%$ and similar additions
are seen in B--V ranging from 1.15 to 1.55.  We now have data that are at 
least uniformly representative across a wide range of temperatures and 
luminosities; however, the sample is still weighted towards observing more 
stars in the most metal--poor population.  Since the new observations 
preferentially target stars with metallicities in the range 
--1.50$\la$[Fe/H]$\la$--0.50, the increased H$^{\rm -}$ opacity and line 
blocking with increasing metallicity causes these stars to have lower flux in 
the spectral regions of interest than their more metal--poor counterparts.  As 
a result, stars observed in progressively more metal--rich branches are, on 
average, more evolved with our magnitude cutoff.

\begin{figure}
\epsscale{1.00}
\plotone{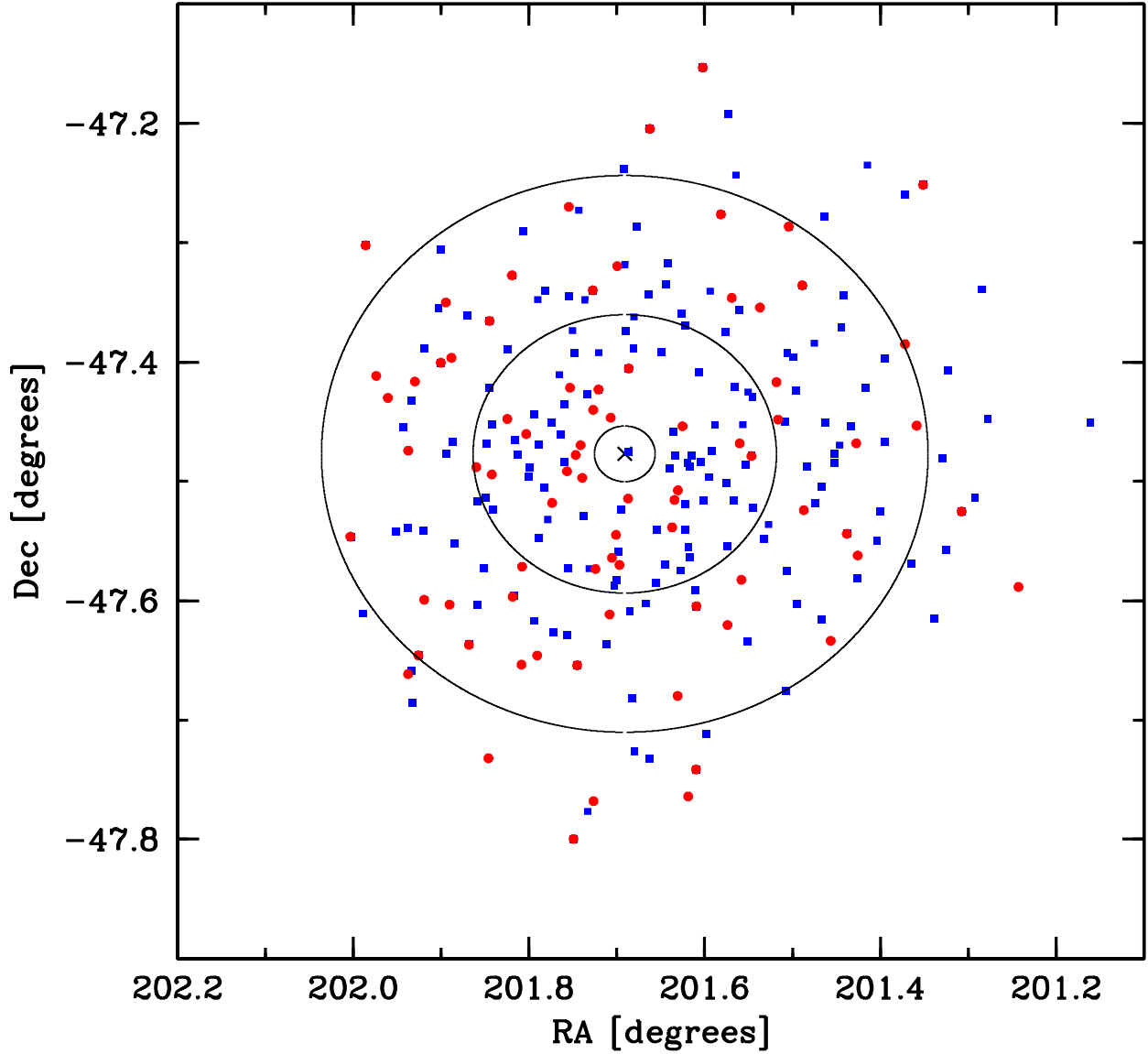}
\caption{Program stars are shown in terms of position in the field.  The 
symbols are the same as in Figure \ref{f1}.  The cross indicates the field
center at 201.691$\degr$, --47.4769$\degr$ (J2000) 
(13$^{\rm h}$26$^{\rm m}$45.9$^{\rm s}$, --47$\degr$28$\arcmin$37.0$\arcsec$) 
and the ellipses indicate 1, 5, and 10 times the core radius of 1.40$\arcmin$.}
\label{f3}
\end{figure}

\begin{figure}
\epsscale{1.00}
\plotone{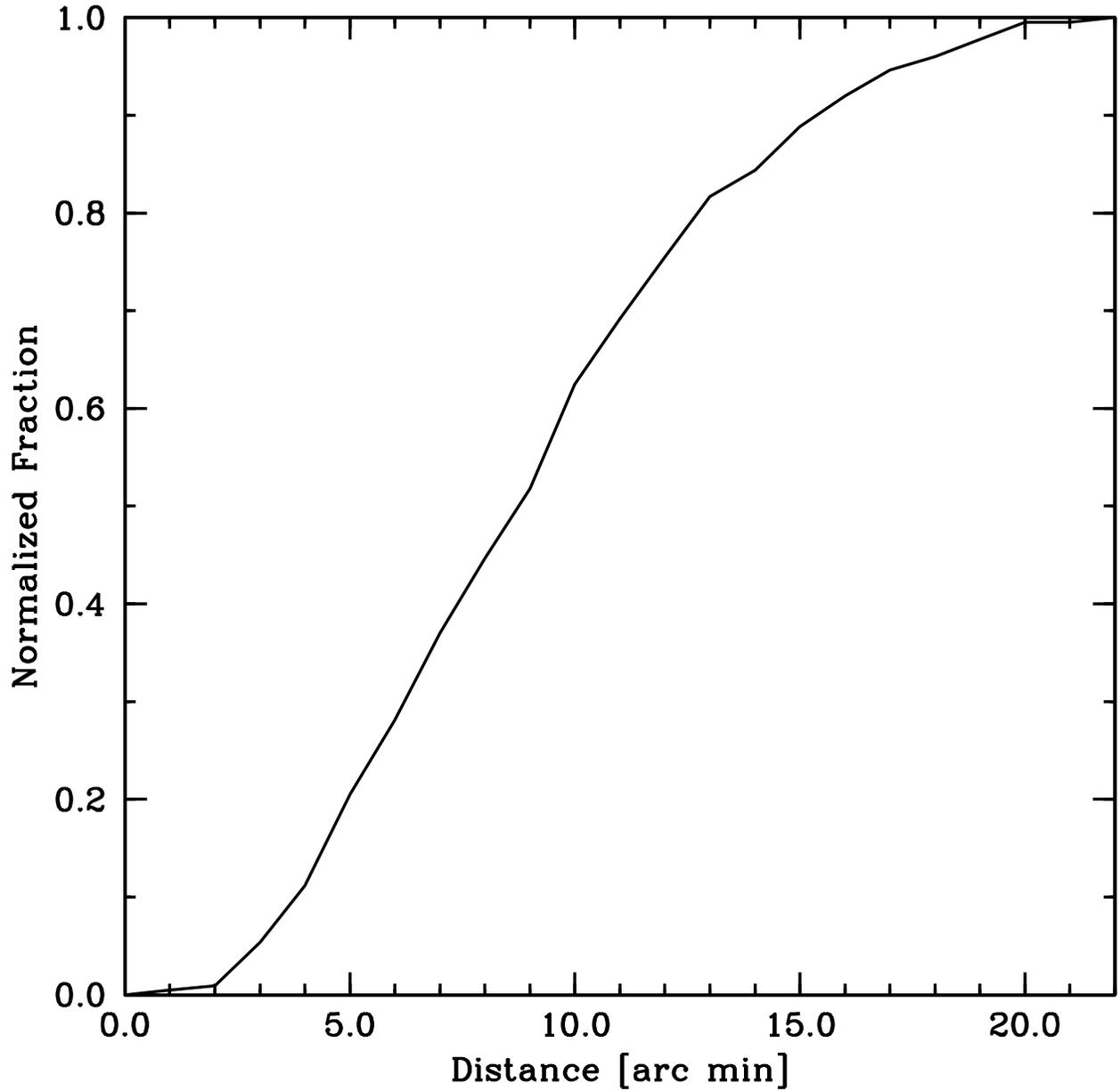}
\caption{Normalized cumulative distribution for our combined sample as a 
function of distance from the cluster center.  This plot shows the fraction
of our total sample observed inside a given radius.  The cluster center 
reference is the same as in Figure \ref{f3}.}
\label{f4}
\end{figure}

There is some evidence for the presence of a radial metallicity gradient in 
the cluster (Norris et al. 1996, 1997; Suntzeff \& Kraft 1996; Hilker \& 
Richtler 2000; Pancino et al. 2000; Rey et al. 2004; Johnson et al. 2008), and
it is important to observe stars at various radii to measure the true 
metallicity distribution.  In Figures \ref{f3} and \ref{f4} we plot the 
positions of our program stars and show a normalized cumulative distribution 
as a function of distance from the cluster center, defined by van Leeuwen et 
al. (2000) as 13$^{\rm h}$26$^{\rm m}$45.9$^{\rm s}$, 
--47$\degr$28$\arcmin$37.0$\arcsec$ (J2000).  Both figures indicate our
combined sample from this study and Johnson et al. (2008) mostly covers stars
between $\sim$5--15$\arcmin$ from the center, which is equivalent to roughly
3.5 to 10.5 core radii where the core radius is 1.40$\arcmin$ (Harris 1996; 
rev. 2003 February).  Fiber positioning limitations and increasing stellar 
densities near the cluster core prevent us from obtaining copious observations 
inside $\sim$1--2$\arcmin$ from the center, but we have observed nearly
30--40$\%$ of all bright giants inside 10--20$\arcmin$.

Data reductions were carried out using various tasks provided in standard
IRAF\footnote{IRAF is distributed by the National Optical Astronomy 
Observatories, which are operated by the Association of Universities for 
Research in Astronomy, Inc., under cooperative agreement with the National 
Science Foundation.} packages.  We used \emph{ccdproc} to trim the overscan
region and apply the bias level corrections.  Flat--field corrections, 
ThAr lamp wavelength calibrations, cosmic ray removal, subtraction of scattered
light and sky spectra, and extraction of the one--dimensional spectra were 
performed using the \emph{dohydra} package.  The resultant spectra were then
corrected for telluric contamination, continuum flattened, and combined.

\section{Analysis}

We have derived abundances for nine different elements using local 
thermodynamic equilibrium (LTE) equivalent width and spectrum synthesis 
analyses in the combined spectral regions of 6000--6250~\AA\ and 
6530--6800~\AA.  Spectrum synthesis was used for determining all Al abundances
because of the potential for CN contamination in metal--rich and CN--strong 
stars.  Model atmosphere parameters including effective temperatures 
(T$_{\rm eff}$), surface gravities (log g), and microturbulence (V$_{\rm t}$) 
were estimated based on published photometry and the empirical 
V$_{\rm t}$--T$_{\rm eff}$ relation given in Johnson et al. (2008).

\subsection{Model Stellar Atmospheres}

Effective temperatures were estimated via empirical calibrations of V and 2MASS
photometry based on the infrared flux method (Blackwell \& Shallis 1977).  To
improve accuracy, we averaged the T$_{\rm eff}$ values obtained through the 
color--temperature relations of Alonso et al. (1999; 2001) and Ram{\'{\i}}rez 
\& Mel{\'e}ndez (2005) for V--J, V--H, and V--K color indices.  The 
photometry was corrected for interstellar reddening and extinction using 
the corrections recommended by Harris (1996) of E(B--V)=0.12 and McCall (2004)
for E(V--J)/E(B--V)=2.25, E(V--H)/E(B--V)=2.55, and E(V--K)/E(B--V)=2.70.  
Evidence for differential reddening is mainly concentrated near the core 
(Calamida et al. 2005; van Loon et al. 2007), but the well defined evolutionary
sequences seen in Villanova et al. (2007) suggest significant differential 
reddening is unlikely.  Therefore, we have applied a uniform reddening 
correction to all stars.  The temperatures derived from each color index are 
in very good agreement with an average offset of 21 K ($\sigma$=6 K).  Our 
adopted T$_{\rm eff}$ values are probably accurate to within $\pm$50 K, and 
are consistent with the star--to--star scatter seen in the calibrations of 
both studies.  Plotting Fe abundance versus excitation potential did not reveal
any trends and our adopted photometric temperatures satisfied excitation 
equilibrium. 

Surface gravity was determined by T$_{\rm eff}$ and absolute bolometric
magnitude (M$_{\rm bol}$) through the standard relation,
\begin{equation}
log(g_{*})=0.40(M_{bol.}-M_{bol.\sun})+log(g_{\sun})+4(log(T/T_{\sun}))+
log(M/M_{\sun}).
\end{equation}
We applied the bolometric correction to M$_{\rm V}$ from Alonso et al. (1999)
and used a distance modulus of (m--M)$_{\rm V}$=13.7 (van de Ven et al. 2006).
As mentioned in $\S$1, an age spread of $\sim$1--4 Gyr is likely present in the 
cluster, but the difference in mass between the oldest and youngest stars is
only of order $\sim$0.05 M$_{\sun}$, which is negligible for surface gravity 
determinations.  Norris et al. (1996) argue that 20--40$\%$ of stars on the 
RGB may be AGB stars with M$\sim$0.60 M$_{\sun}$, but this would only lead to 
abundance uncertainties of order 0.10 dex for species residing in the dominant 
ionization state (e.g., Fe II).  Similar surface gravity and abundance effects 
may be expected for He--rich stars, which have slightly lower RGB masses due to
their shorter lifetimes compared to He--normal stars (e.g., Newsham \& 
Terndrup 2007).

\begin{figure}
\epsscale{0.90}
\plotone{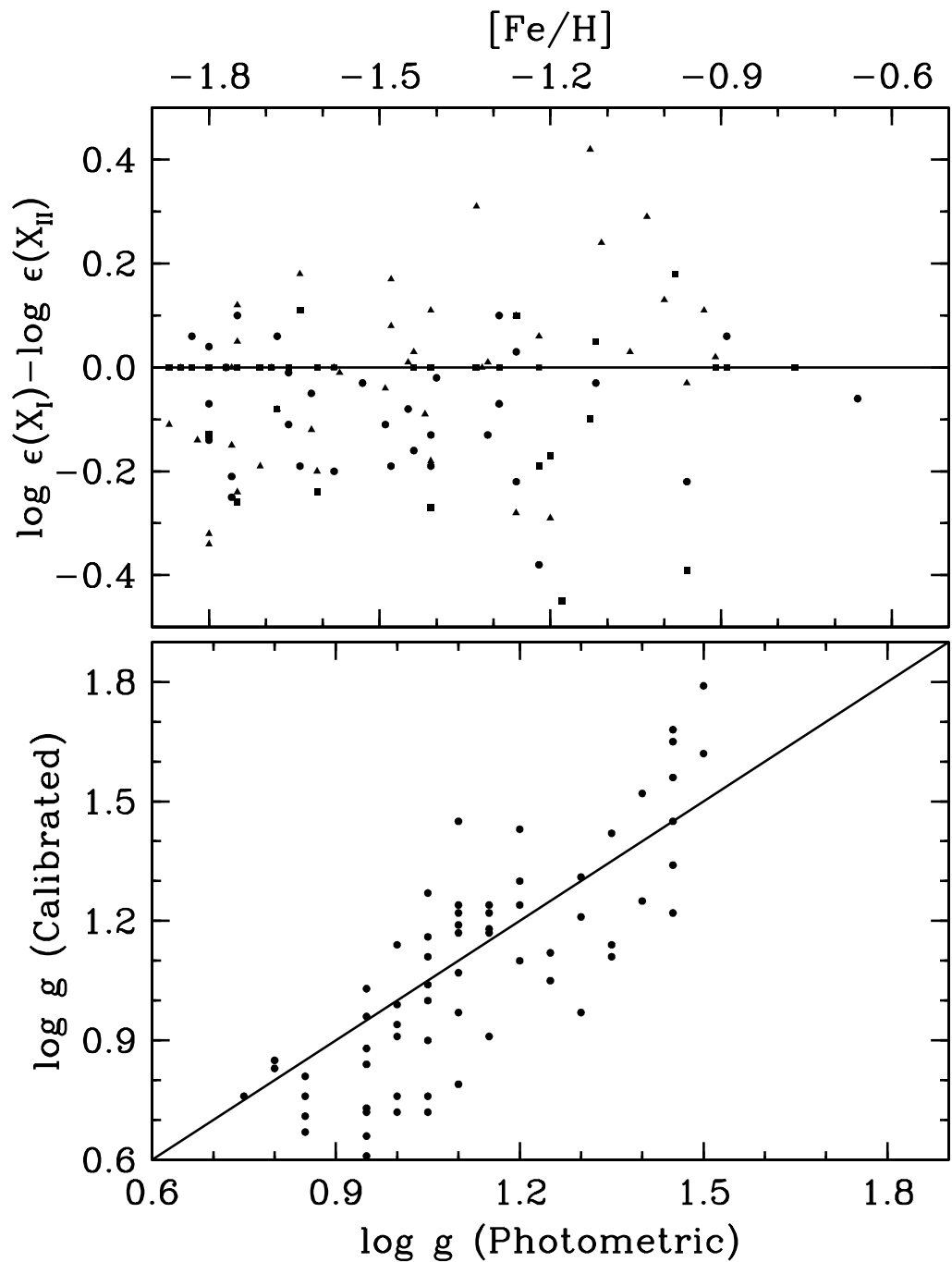}
\caption{The top panel shows a plot of the difference in abundance as derived
from both neutral and singly ionized species as a function of [Fe/H].  The
filled circles represent Fe, the filled boxes are Sc, and the filled triangles
are Ti.  The bottom panel shows the log g values adopted from photometry
versus the calibrated T$_{\rm eff}$--log g relation from Ku{\v c}inskas et al.
(2006).  In both panels the straight line indicates perfect agreement.}
\label{f5}
\end{figure}

Since we only had a limited number of singly ionized lines available for
analysis relative to the number of neutral lines, we relied on the photometric
surface gravity estimate instead of ionization equilibrium.  In the top panel
of Figure \ref{f5}, we show the differences in derived abundance for neutral
and singly ionized species of Fe, Sc, and Ti.  For Fe, the average offset
between log $\epsilon$(Fe I) and log $\epsilon$(Fe II) is --0.09
($\sigma$=0.12), but this is based on a highly discrepant number of lines
between the two species and thus may not accurately reflect a systematically
low gravity.  Sc shares a similar pattern with an average difference of --0.16
dex ($\sigma$=0.22), but these are based on only one line a piece for each
species and reliable hyperfine structure information for these transitions is
sparse.  Ti I and II lines give an average difference of --0.01 dex 
($\sigma$=0.18).  Overall, the difference between abundances derived from both 
species is --0.07 dex ($\sigma$=0.17), which is comparable to measurement and 
model uncertainties.  In the bottom panel of Figure \ref{f5}, we show 
photometric log g values compared with estimates based on spectroscopic 
gravity calibrations of globular clusters provided by Ku{\v c}inskas et al. 
(2006).  The average offset between the two systems is +0.04 dex 
($\sigma$=0.17), and is comparable to the uncertainty found by examining 
ionization equilibrium.  This leads us to believe our surface gravity 
estimates are not in serious error.

We obtained a rough estimate of [Fe/H] using the [Ca/H] calibration based on 
V and B--V given in van Leeuwen et al. (2000; their equation 15) and assumed 
[Ca/Fe]$\sim$+0.30.  Likewise, an initial microturbulence value was calculated
from the V$_{\rm t}$--T$_{\rm eff}$ relation given in Johnson et al. (2008)
for luminous giants.  Our determined T$_{\rm eff}$, log g, [Fe/H], and 
V$_{\rm t}$ values were used to generate model atmospheres (without convective 
overshoot) via interpolation within the ATLAS9\footnote{The model atmosphere 
grids can be downloaded from http://cfaku5.cfa.harvard.edu/grids.html.} grid 
(Castelli et al. 1997).  We iteratively adjusted the microturbulence of the 
model until Fe abundances were independent of line strength following the 
method described in Magain (1984).  Lastly, the model's metallicity was
adjusted to match the derived Fe abundance of each star.  A complete list of
our adopted model atmosphere parameters along with star identifiers, published
photometry, membership probabilities, and S/N estimates for each spectrum are 
provided in Table 1.

\subsection{Derivation of Abundances}

\subsubsection{Equivalent Width Analysis}

For all elements except Al, final abundances were determined using equivalent
width analyses and the \emph{abfind} driver in the LTE line analysis code MOOG
(Sneden 1973).  Equivalent widths were measured by interactively fitting 
multiple Gaussians to isolated and blended line profiles.  Suitable lines were 
identified using the solar and Arcturus atlases\footnote{The atlases can be 
downloaded from the NOAO Digital Library at 
http://www.noao.edu/dpp/library.html.}.  Our adopted log gf values were 
determined by measuring equivalent widths in the solar atlas and then modified 
until all lines yielded abundances consistent with the photospheric values 
given in Anders \& Grevesse (1989).  A summary of our line list and measured 
equivalent widths is given in Tables 2a--2f and the final abundances are in 
Table 3.  Note that all abundance ratios in Table 3 are relative to 
[Fe/H]$_{\rm avg.}$, which is the average of [Fe I/H] and [Fe II/H] or just 
[Fe I/H] if Fe II lines are not available.  We did not determine [X/Fe] ratios 
by matching ionization states because many stars did not have reliable Fe II 
transitions, our typical measured [Fe I/H] values are based on more than 25 
lines versus 1 or 2 for [Fe II/H], and $\langle$[Fe I/H]--[Fe II/H]$\rangle$ 
is approximately equal to the line--to--line dispersion seen in our [Fe I/H] 
determinations.  Giving [X/Fe] ratios relative to [Fe/H]$_{\rm avg.}$ is an 
attempt to minimize the effects of overionization.

Many line profiles for the odd--Z Fe--peak and neutron capture elements suffer
from hyperfine splitting, but the necessary atomic data for several of these 
transitions are not currently available in the literature.  A standard 
equivalent width analysis will produce an overabundance if this effect is not 
properly taken into account.  Since the error caused by hyperfine splitting 
increases with line strength, we do not expect our abundances, which are based 
on unsaturated and generally weak lines, to be strongly affected.  

The elements here that may be affected by hyperfine splitting are Sc, La, and 
Eu.  Our Sc abundances are based on the 6210.67~\AA\ Sc I line and 
6604.60~\AA\ Sc II line.  While hyperfine structure estimates have been 
produced for Sc II (Prochaska \& McWilliam 2000), there is no available 
information for the Sc I line and neither of these transitions is included in 
Zhang et al. (2008).  Therefore, we have not applied the correction to Sc II, 
but the offset is probably not too large given that the average equivalent 
width for the Sc I/II lines is $\sim$60 m\AA.  Similarly, no hyperfine data 
exist for the 6774.27~\AA\ La II line and therefore we accept the derived 
abundances at face value.  Europium is slightly more complicated because, in 
addition to hyperfine splitting, it has two stable, naturally occurring 
isotopes (\iso{151}{Eu} and \iso{153}{Eu}) that are present in nearly equal 
proportions.  For all Eu abundances, we have applied a hyperfine and isotopic 
line list provided by C. Sneden (2006, private communication).  Lacking 
\emph{a priori} knowledge of the r--/s--process contributions for La and Eu in 
$\omega$ Cen, we have assumed a solar mix such that the ratio for La is 
25/75$\%$ (Sneden et al. 2008) and for Eu 97/3$\%$ (Sneden et al. 1996), 
respectively.

\subsubsection{Spectrum Synthesis Analysis}

\begin{figure}
\epsscale{0.85}
\plotone{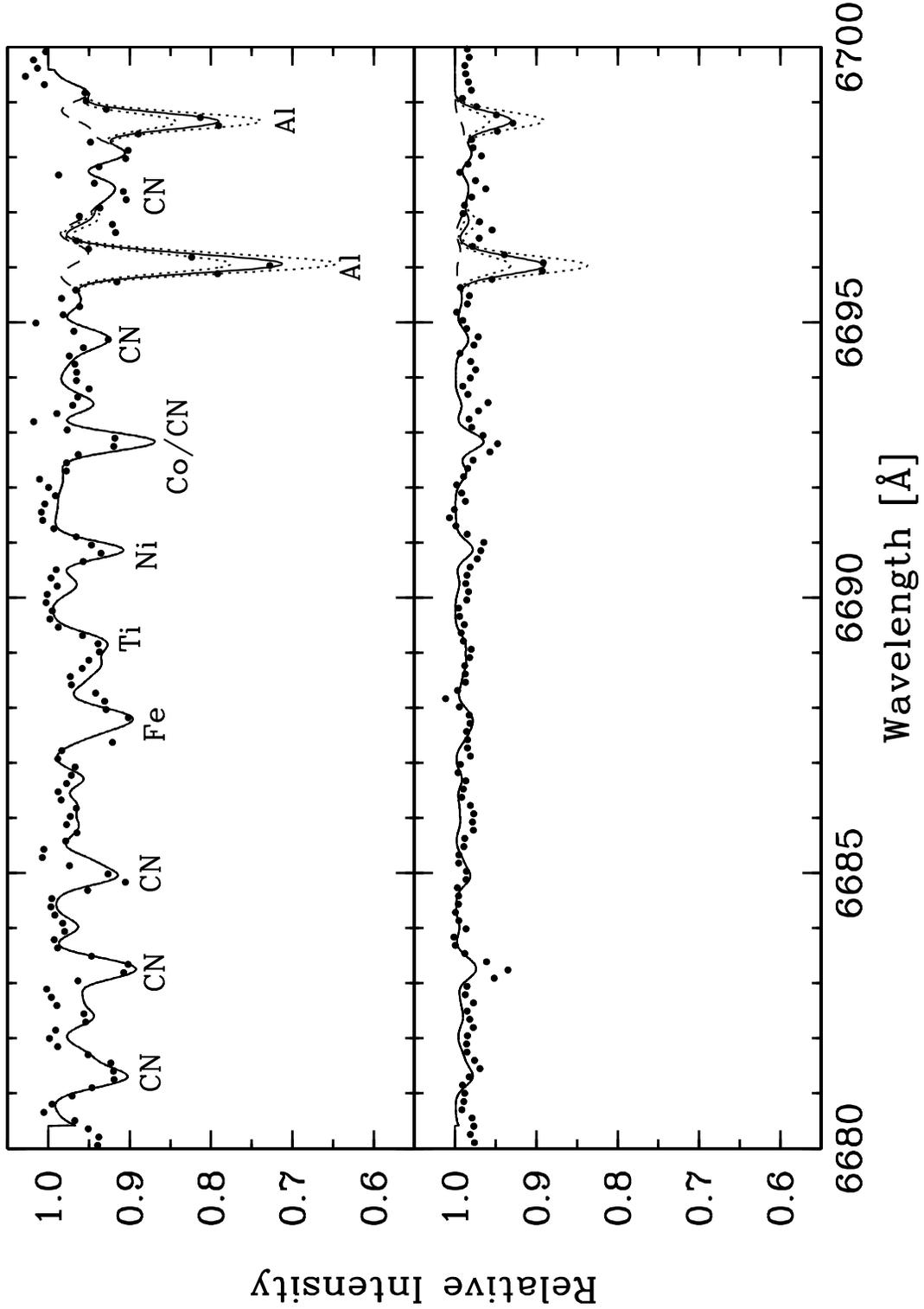}
\caption{Sample spectrum syntheses are shown for two stars of varying 
T$_{\rm eff}$, [Fe/H], and CN strength, but similar [Al/Fe] ratios.  The 
relative intensity scales are the same in both figures.  The solid line shows
the best fit to the observed spectrum, the dotted lines illustrate deviations
$\pm$0.30 dex, and the dashed line indicates how the spectrum would appear if
Al were absent.}
\label{f6}
\end{figure}

While all other abundances were determined using equivalent width analyses,
we derived Al abundances via spectrum synthesis because of the non--trivial
contamination from CN lines seen in many of the cooler, more metal--rich stars.
For consistency, spectrum synthesis was performed even in cases where CN
contamination was not an issue.  In Figure \ref{f6} we show two sample 
syntheses for a case with strong (top panel) and weak (bottom panel) CN lines
in order to illustrate the line blanketing effects from molecular absorption.  
For stars where the CN lines were present, we found that a straight--forward 
equivalent width analysis led to an over estimate of log $\epsilon$(Al) by as 
much as 0.1--0.2 dex compared to spectrum synthesis, but the two methods 
agreed to within $<$0.1 dex in spectra without strong CN features.  The Al 
lines are designated by ``synth" in Tables 2a--2f.

We created the molecular line list by combining the Kurucz online 
database\footnote{The Kurucz line lists can be accessed 
at: http://kurucz.harvard.edu/linelists.html.} with one provided by B. Plez
(private communication, 2007; see also Hill et al. 2002).  Since the C, N, and 
\iso{12}{C}/\iso{13}{C} abundances are unknown, we fixed [C/Fe]=--0.5 and 
\iso{12}{C}/\iso{13}{C}=4, which are consistent with Norris \& Da Costa 
(1995) and Smith et al. (2002) for $\omega$ Cen giants.  Without accurate C, 
N, and O data, it is impossible to constrain the molecular equilibrium 
equations and derive true C and/or N abundances for CN.  Therefore, we treated 
the nitrogen abundance as a free parameter and adjusted it to obtain a best 
fit to the CN lines.  Other metal lines surrounding the Al doublet have 
excitation potentials $\ga$5 eV and are not important contributors in these 
cool stars.

\subsubsection{Abundance Comparison to Other Studies}

Although $\omega$ Cen has been the subject of many spectroscopic studies, here
we restrict comparison to those measuring abundances using moderately high
resolution (R$\ga$15,000) spectroscopy.  The only two previous works for which
we have several stars in common are Norris \& Da Costa (1995) and Johnson
et al. (2008).  In the case of Norris \& Da Costa, the average difference in
measured [Fe/H] for the 7 common stars is --0.02 dex ($\sigma$=0.05), in the
sense present minus Norris \& Da Costa.  The results are similar for most of
the other elements with average differences of order $\pm$0.10 dex 
($\sigma$$\sim$0.15), and La is the only exception with an average offset of
$+$0.34 ($\sigma$=0.12).  This discrepancy is likely due to the difficulty in
obtaining accurate La abundances.  In comparison to Johnson et al. (2008), the 
difference in [Fe/H] based on 21 stars in common is --0.10 dex ($\sigma$=0.05)
and 0.00 dex ($\sigma$=0.22) for [Al/Fe].  These results are consistent with
the small deviations in adopted atmospheric parameters found by Johnson et al.
(2008; see their Figures 8--9) comparing that study to other spectroscopic 
surveys in the literature.

\subsection{Abundance Sensitivity to Model Atmosphere Parameters}

In Table 4, we show the results of our tests regarding abundance sensitivity
to uncertainties in adopted model atmosphere parameters for all elements 
studied here.  We examined how the various log $\epsilon$(X) values changed 
when altering T$_{\rm eff}$$\pm$100 K, log g$\pm$0.30 cm s$^{\rm -2}$, 
[M/H]$\pm$0.30 dex, and V$_{\rm t}$$\pm$0.30 km s$^{\rm -1}$.  In general, we 
find that the neutral species tend to be more strongly affected by changes in 
temperature, but the singly ionized species are influenced by surface gravity
changes because of their dependence on electron pressure.  However, abundances
taken from singly ionized transitions tend to have a larger dependence on
T$_{\rm eff}$ with increasing metallicity while the effects on neutral lines
are mitigated.  Similarly, only the ionized species have a significant 
dependence on the model atmosphere's overall metallicity because their 
line--to--continuous opacity ratios are more sensitive to changes in the 
H$^{\rm -}$ abundance.  For stars with [Fe/H]$\la$--1, microturbulence 
uncertainties have very little influence on the derived abundance because 
the lines are weak and lie further down the linear portion of the 
curve of growth, but even in the most metal--rich stars the effects are 
typically no larger than $\pm$0.10--0.15 dex.  The lanthanum line is an 
exception because the more metal--rich $\omega$ Cen stars are very s--process
rich and thus the La II lines typically have equivalent widths $\ga$75 m\AA.
Although each element has a slightly different dependence on these physical
parameters, the important point is that the element--to--iron ratio should be 
mostly invariant.  Instead, only the log $\epsilon$(X) values should be 
sensitive to model parameter variations.

As mentioned in $\S$1, it has been argued that several of the intermediate and 
perhaps metal--rich stars in this cluster may have strong He enhancements 
extending as large as Y$\sim$0.38.  We do not expect our analysis to be 
severely altered (see Girardi et al. 2007) and the [X/Fe] ratios should be 
mostly independent of the adopted He abundance; however, [Fe/H] may be 
systematically higher in the He--rich stars.  To test the effects of He 
enhancement, we created synthetic spectra using He--normal (Y=0.27) and 
He--rich (Y=0.35) ATLAS9 models of comparable temperature and metallicity to 
our stars.  We found the line strength differences between the two sets to be 
much less than 1$\%$, with the He--rich model producing stronger lines because 
of decreased continuous H$^{\rm -}$ opacity.  This result is consistent with 
Piotto et al. (2005) and leads us to believe our abundances are robust against 
possible He variations.

\section{RESULTS}

\subsection{Light elements: Na \& Al}

\begin{figure}
\epsscale{0.90}
\plotone{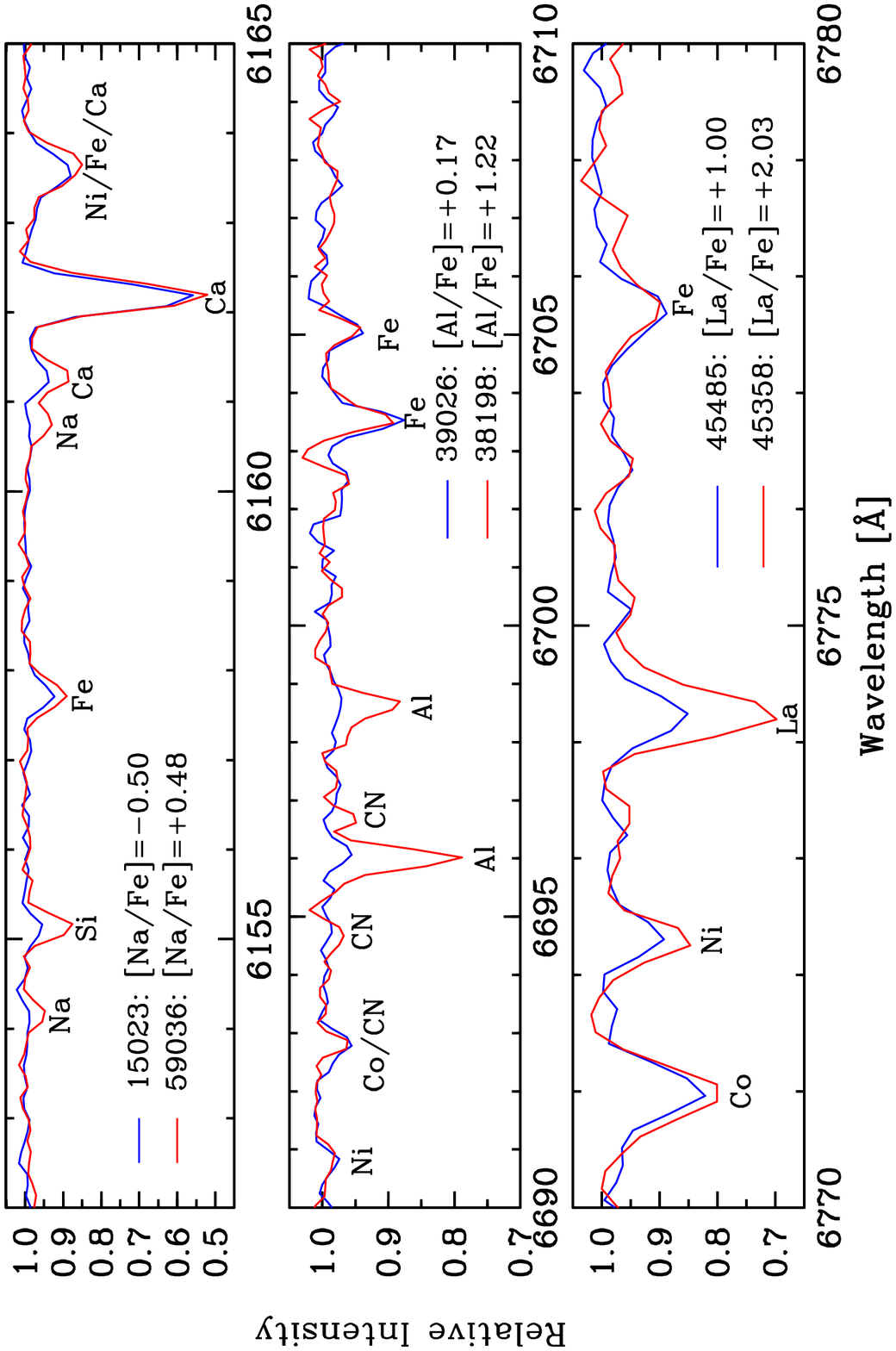}
\caption{Sample spectra are shown in three different wavelength regions to
highlight the line strength differences seen in Na, Al, and La.  Each panel
contains stars of roughly the same T$_{\rm eff}$ and [Fe/H].  Note the
differences seen in both the Si and Ca features compared to Fe in the top
panel.}
\label{f7}
\end{figure}

The odd--Z elements Na and Al are particularly important because they are 
among the heaviest elements thought to be produced via proton--capture 
nucleosynthesis in low and intermediate mass stars.  This makes them useful
probes for deciphering which processes, in addition to Type II SN production, 
may have been dominant during various epochs of star formation.  Previous
large sample, high resolution spectroscopic studies of $\omega$ Cen giants 
(e.g., Norris \& Da Costa 1995; Smith et al. 2000; Johnson et al. 2008) have
shown that Na and Al (in addition to C, N, O, and Mg) exhibit very large
star--to--star [X/Fe] variations while preserving the Na--Al correlation seen
in other Galactic globular clusters.  The top two panels of Figure \ref{f7}
illustrate this point by demonstrating the stark contrast in Na and Al line 
strengths for stars with similar temperatures and metallicities.  Since we 
can compare stars of similar evolutionary state and metallicity, we can 
safely assume departures from LTE are not the cause of the observed abundance
variations.  No NLTE corrections have been applied to our Na and Al results 
because there are no ``standard" values available in the literature and those 
that are available disagree in magnitude and sign.  However, Na and Al 
abundances determined from the non--resonance, subordinate transitions used 
here typically have corrections of order $\la$0.20 dex for stars with
--2.5$<$[Fe/H]$<$--0.5 (e.g., Gratton et al. 1999; Gehren et al. 2004).

First considering our Na results, we find that there is a general increase in
$\langle$[Na/Fe]$\rangle$ as a function of increasing metallicity accompanied
by a decrease in the star--to--star scatter.  The dominant metallicity group 
of stars (--1.8$\la$[Fe/H]$\la$--1.6) have $\langle$[Na/Fe]$\rangle$=+0.03 
($\sigma$=0.32) with a full range of 1.29 dex while the next population of 
stars (--1.5$\la$[Fe/H]$\la$--1.3) have $\langle$[Na/Fe]$\rangle$=+0.20 
($\sigma$=0.21) and a full range of 0.67 dex, which is smaller by about a 
factor of 4.  However, there is a noticeable change in the distribution of 
[Na/Fe] for RGB stars in the higher metallicity populations.  At 
[Fe/H]$\ga$--1.2, 95$\%$ (18/19) of the stars are very Na--rich with
$\langle$[Na/Fe]$\rangle$=+0.86 ($\sigma$=0.12).  The strong enrichment of
this population is in agreement with Norris \& Da Costa (1995) who find that 
at least 75$\%$ (6/8) of stars in that metallicity range are Na--rich and at 
least 50$\%$ are O--poor.  A two--sample Kolmogorov--Smirnov (K--S) test (Press 
et al. 1992) shows that the population of stars with [Fe/H]$<$--1.2 is drawn 
from a different parent population than the [Fe/H]$>$--1.2 group at the 99$\%$ 
level.

\begin{figure}
\epsscale{1.00}
\plotone{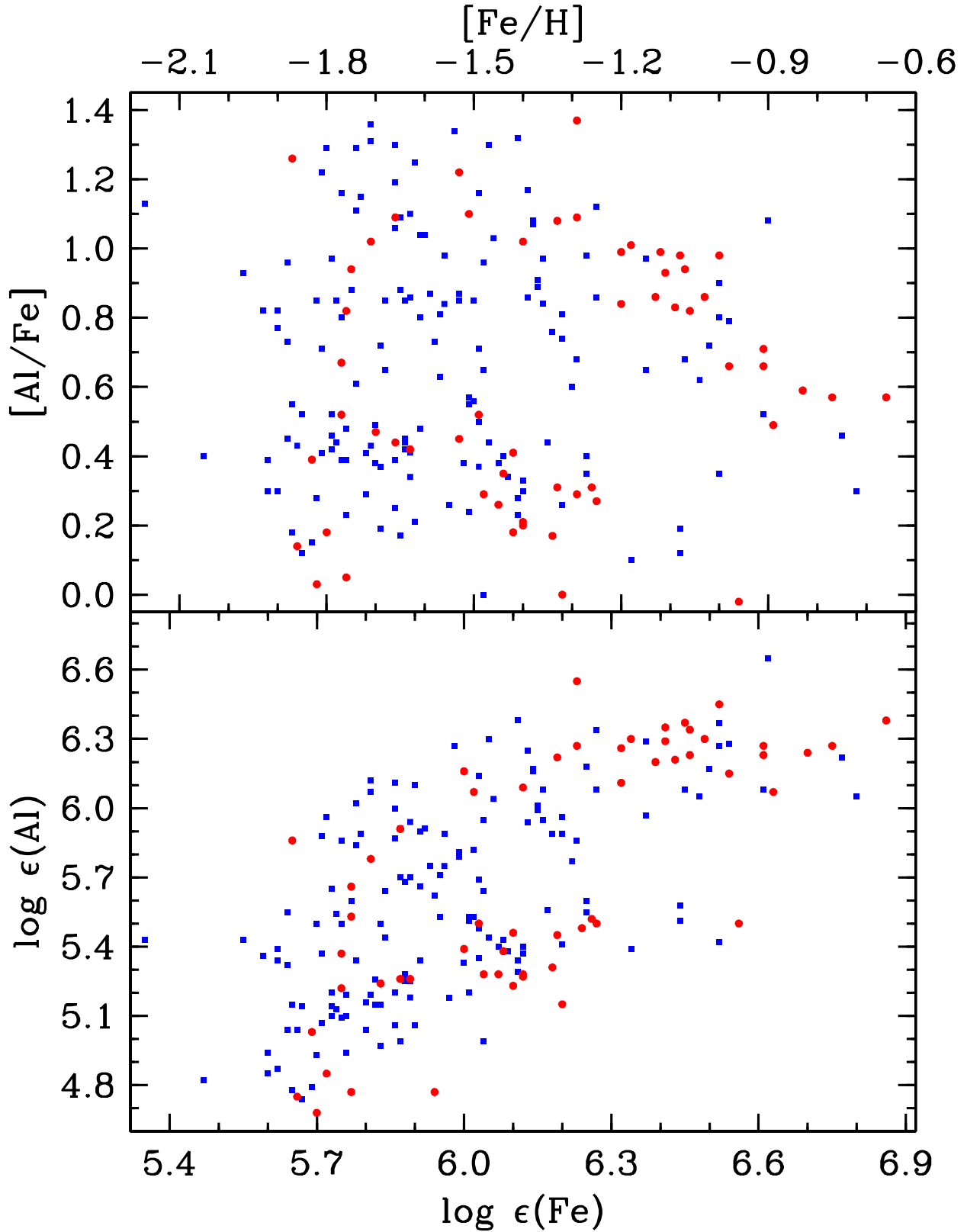}
\caption{The top panel shows [Al/Fe] plotted as a function of [Fe/H] and the
bottom panel shows log $\epsilon$(Al) plotted as a function of log
$\epsilon$(Fe).  The symbols are the same as those in Figure \ref{f1}.}
\label{f8}
\end{figure}

By combining our current data with that of Johnson et al. (2008), we have a
homogeneous set of [Al/Fe] abundances determined for more than 200 RGB stars.
In Figure \ref{f8} we show the results of our combined samples for [Al/Fe] and 
log $\epsilon$(Al) as a function of metallicity.  Although the sample sizes
between Na and Al differ by a factor of 3.5, some key differences standout in 
the Al data set:  (1) there appear to be 2 or 3 different populations of stars,
(2) the star--to--star dispersion stays mostly constant until 
[Fe/H]$\sim$--1.2, and (3) stars with [Fe/H]$\ga$--1.2 show a roughly constant 
log $\epsilon$(Al)$\approx$6.22 ($\sigma$=0.18) as a function of increasing 
[Fe/H].  However, there are some interesting similarities: (1) the Al data show
a clear change in the abundance pattern for stars with [Fe/H]$\ga$--1.2, 
(2) the metal--rich RGB stars are predominantly Al--rich, and (3) 
log $\epsilon$(Na)$_{\rm max}$$\approx$log $\epsilon$(Al)$_{\rm max}$.  It
should be noted that despite the large abundance scatter, the Na--Al 
correlation is present in our data.

We define the three different Al populations as those having
[Al/Fe]$<$+0.60, +0.60$\le$[Al/Fe]$<$+1.0, and [Al/Fe]$\ge$+1.0.  First 
considering only stars with [Fe/H]$<$--1.2, the subpopulations break down into
$\langle$[Al/Fe]$\rangle$=+0.34 ($\sigma$=0.14), 
$\langle$[Al/Fe]$\rangle$=+0.82 ($\sigma$=0.10), and
$\langle$[Al/Fe]$\rangle$=+1.17 ($\sigma$=0.11), respectively.  These represent
50$\%$ (83/166), 30$\%$ (49/166), and 20$\%$ (34/166) of the cluster stars in
this metallicity regime.  Extending this break down to the entire sample gives
a similar distribution of 48$\%$ (96/202), 34$\%$ (69/202), and 18$\%$ 
(37/202), respectively.  This distribution is perhaps tied to the 
``primordial", ``intermediate", and ``extreme" populations of the O--Na 
anticorrelation (Carretta et al. 2008).  However, only the intermediate Al
subpopulation is present at all metallicities.  The very enhanced Al stars 
([Al/Fe]$>$+1) are only found at [Fe/H]$\la$--1.2, and the sequence of low--Al 
stars ([Al/Fe]$<$+0.60) essentially terminates at about the same metallicity 
cut--off (this is particularly evident  in the bottom panel of 
Figure \ref{f8}).  A two--sample K--S test confirms the same result as the Na 
case, which is that the [Al/Fe] distribution for stars with [Fe/H]$>$--1.2 and 
[Fe/H]$<$--1.2 are different at the 94$\%$ level.

These data suggest that $\omega$ Cen's metal--rich populations may be 
significantly more chemically homogeneous than the metal--poor (and presumably
older) populations, but the gas from which the metal--rich stars formed was
enhanced in light elements at a level beyond what is thought to be possible 
from Type II SNe (e.g., Woosley \& Weaver 1995; Chieffi \& Limongi 2004).  
Evidently, at high metallicity it is possible to produce Na in greater 
quantities than Al.

\subsection{$\alpha$ elements}

The $\alpha$ elements are often used to gauge the relative contributions from
Type II SNe, which are efficient producers of $\alpha$ elements, and Type Ia
SNe, which produce mostly Fe--peak elements.  Predicted stellar yields from 
core collapse SNe weighted by a Salpeter initial mass function (IMF; x=1.35) 
produce [$\alpha$/Fe]$\sim$+0.30 to +0.50 across a broad range of metallicities
(e.g., Chieffi \& Limongi 2004).  Therefore, values of [$\alpha$/Fe]$\sim$+0.10
or less suggest Type Ia SNe have contributed some portion of the Fe--peak 
elements.  The most commonly measured $\alpha$ elements are O, Mg, Si, Ca, and 
Ti; however, Ti is more complicated because it has multiple production sources.
In globular clusters, a large portion of the stars have had their O and Mg 
abundances altered by proton--capture nucleosynthesis and therefore these 
elements cannot be treated as ``pure" $\alpha$ elements.  This restricts 
discussions regarding $\alpha$ enhancement to the heavier elements.  

Previous studies of $\omega$ Cen and other globular clusters have shown nearly
all stars to be $\alpha$ enhanced at [$\alpha$/Fe]$\sim$+0.40 with very small
star--to--star scatter (e.g., see review by Gratton et al. 2004).  Our results 
are consistent with previous work and give $\langle$[Ca/Fe]$\rangle$=+0.36 
($\sigma$=0.09).  Although Ti may straddle being classified as an $\alpha$ or
Fe--peak element, the stars in our sample are mostly Ti--enhanced with 
$\langle$[Ti/Fe]$\rangle$=+0.23 ($\sigma$=0.14).  We do not find any stars to 
be $\alpha$--poor and our lowest derived value is [Ca/Fe]=+0.17, but a handful 
of $\alpha$--poor stars have been found in this cluster (e.g., Norris \& Da 
Costa 1995; Smith et al. 1995, 2000; Pancino et al. 2002).  We do not find any 
trend in [Ca/Fe] with [Fe/H], which is in contrast to the results of Pancino 
et al. (2002) who find the most metal--rich stars to be $\alpha$--poor.  
However, there is real scatter in [$\alpha$/Fe] in this cluster as is evident 
in the Si and Ca line strength variations seen in the top panel of Figure 
\ref{f7}.  In any case, larger sample sizes are required to settle this issue, 
but it seems that the majority of $\omega$ Cen stars are $\alpha$--rich and 
thus Type Ia SNe have not contributed much to the [X/Fe] ratios.  This is a 
significant problem from a chemical evolution standpoint because either the  
ejecta from Type Ia SNe were preferentially lost or their presence was
suppressed despite a several Gyr timespan in star formation.

\subsection{Fe \& Fe--peak elements}

As mentioned above, Fe--peak elements are produced in both Type II and Type Ia
SNe in copious amounts and are the most commonly used tracers of metallicity 
in stars.  These elements are produced in similar conditions during the late 
stages of stellar evolution and as a result often track together as a function
of overall metallicity.  Aside from Fe, the other Fe--peak elements analyzed 
here reproduce this trend with $\langle$[Sc/Fe]$\rangle$=+0.09 ($\sigma$=0.15) 
and $\langle$[Ni/Fe]$\rangle$=--0.04 ($\sigma$=0.09).  In both cases, there
is no trend in [X/Fe] with [Fe/H].  However, since $\omega$ Cen hosts multiple
stellar populations, the behavior of [Fe/H] is not as simple as most 
monometallic globular clusters.

\begin{figure}
\epsscale{1.00}
\plotone{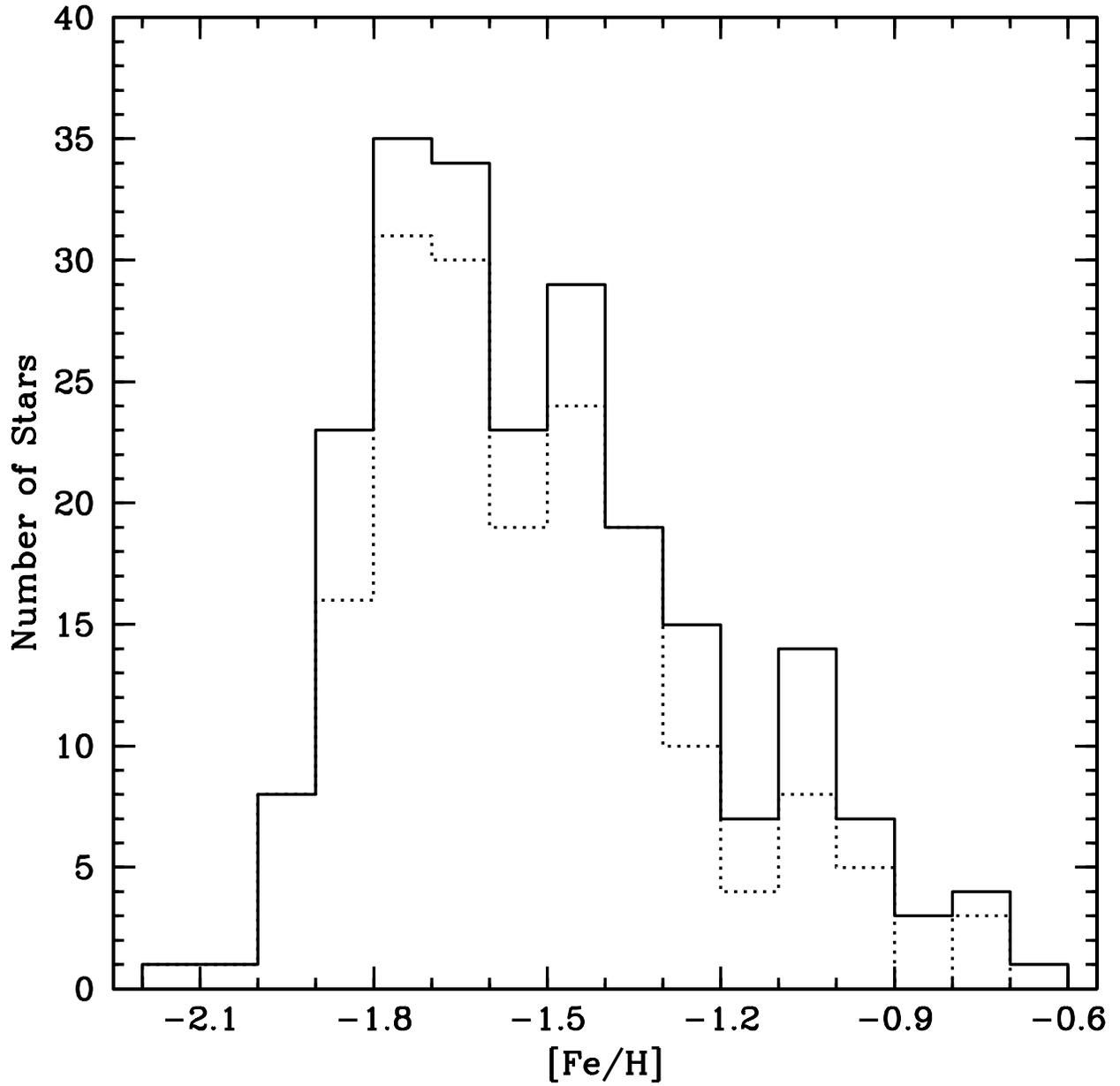}
\caption{Histogram of derived [Fe/H] values for the combined sample of this
study and Johnson et al. (2008) with bin sizes of 0.10 dex.  The dashed line
histogram shows the results from Johnson et al. (2008).}
\label{f9}
\end{figure}

Large sample spectroscopic and photometric observations of $\omega$ Cen (e.g.,
Norris \& Da Costa 1995; Suntzeff \& Kraft 1996; van Leeuwen et al. 2000; Rey 
et al. 2004; Sollima et al. 2005a; Villanova et al. 2007; Johnson et al. 2008)
have shown that the cluster hosts multiple populations of stars 
with almost no stars being more metal--poor than [Fe/H]=--2, more than half
having [Fe/H]$\approx$--1.7, and the rest forming a high metallicity tail 
extending to [Fe/H]$\sim$--0.5.  Again combining our new results with those
from Johnson et al. (2008), we have a homogeneous set of spectroscopically 
determined [Fe/H] abundances for 228 RGB stars.  In Figure \ref{f9} we show
a histogram of our combined sample and our results are consistent with the 
cluster having multiple peaks in the metallicity distribution function at
[Fe/H]=--1.75, --1.45, --1.05, and --0.75.  These peaks constitute roughly
55$\%$, 30$\%$, 10$\%$, and 5$\%$ of our observations, respectively.  The 
percentage of stars contained in each population is nearly identical between
our entire sample and a subset having the highest completion fraction 
(V$\leq$12.5).  This leads us to believe our full sample is representative of
the entire cluster population.  

\begin{figure}
\epsscale{1.00}
\plotone{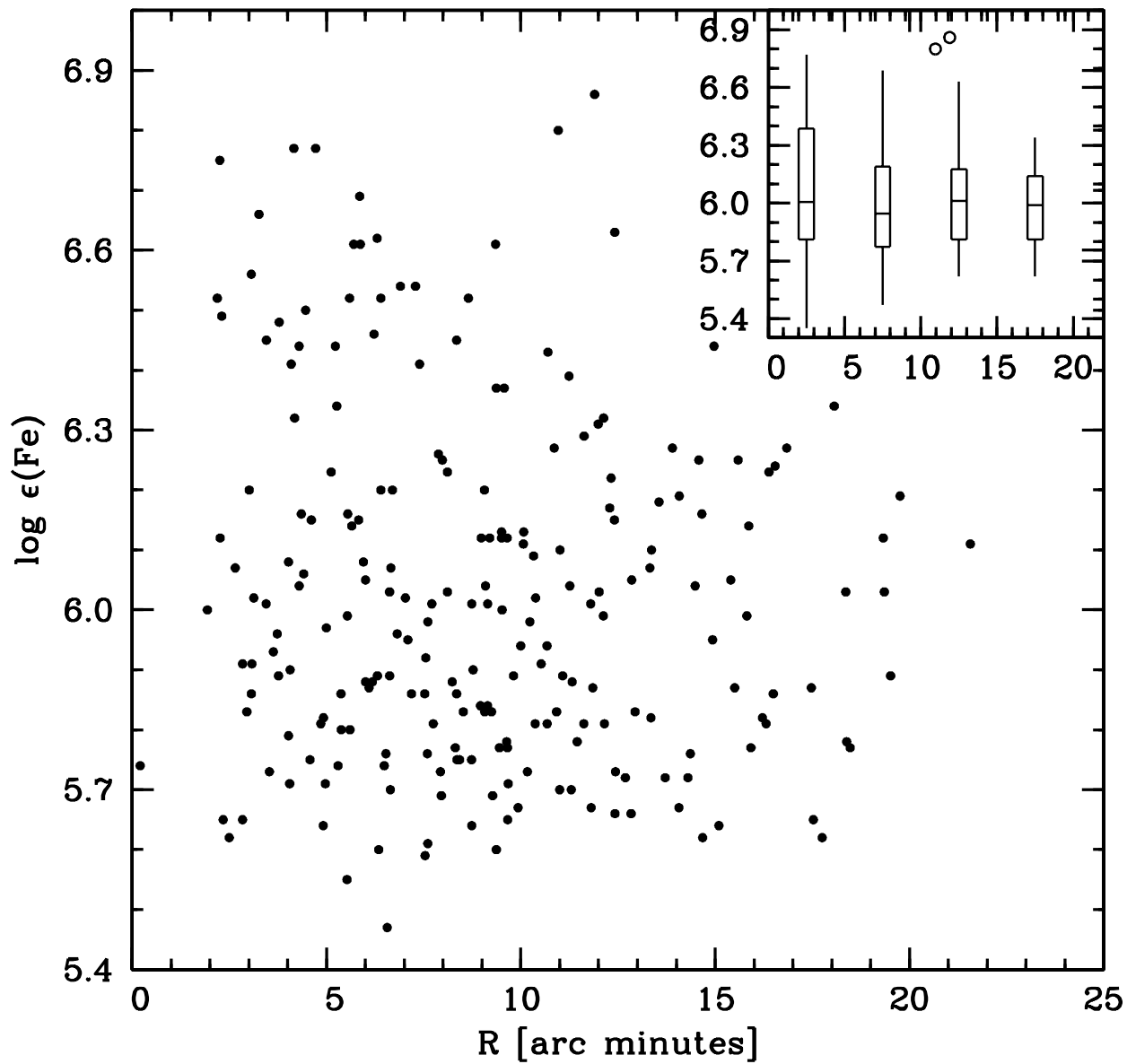}
\caption{Fe is plotted as a function of distance from the cluster center.  The
points show the data from both this study and Johnson et al. (2008).  We have 
averaged the Fe abundances for stars observed in both studies.  The inset plot
shows the mean and quartile distributions in 5$\arcmin$ bins.  The vertical 
lines represent the full data range (except outliers) and open circles 
indicate mild outliers between 1.5 and 3.0 times the interquartile range.}
\label{f10}
\end{figure}

In addition to $\omega$ Cen being chemically diverse, there is some evidence
for a cluster metallicity gradient such that the inner regions contain 
most of the metal--rich stars (e.g., Suntzeff \& Kraft 1996; Norris et al. 1996
Hilker \& Richtler 2000; Pancino et al. 2003; Johnson et al. 2008).  These
results are confirmed in photometric studies (e.g., Rey et al. 2004), which
show that the anomalous, metal--rich RGB (RGB--a) is found only near the 
core of the cluster.  In Figure \ref{f10}, we plot log $\epsilon$(Fe) versus 
distance from the cluster center.  We find that the most metal--rich stars are 
mostly located inside 10$\arcmin$, but the metal--poor stars are located 
uniformly throughout the cluster.  The inset plot in Figure \ref{f10} shows 
that the median metallicity stays constant at about log $\epsilon$(Fe)=6.0 
([Fe/H]$\approx$--1.5) at all radii, but the metallicity interquartile and 
full ranges decrease at large radii.  There has been speculation that, in 
addition to these spatial anomalies, the various cluster populations may 
exhibit unique kinematic signatures as the result of the cluster formation
process (e.g., Norris et al. 1997; Sollima et al. 2005b).  However, recent 
larger sample studies seem to indicate none of $\omega$ Cen's subpopulations 
exhibit rotational, proper motion, or radial velocity anomalies (Reijns et al.
2006; Pancino et al. 2007; Johnson et al. 2008; Bellini et al. 2009).  These
new results seem to negate the merger and background cluster superposition 
scenarios.

\subsection{Neutron--capture elements}

Most elements heavier than the Fe--peak are produced via successive neutron
captures on seed nuclei, but a limited number of these elements have optical
atomic transitions.  In the metallicity regime considered here, Ba and La
are often the primary tracers of the main s--process component and Eu the
primary tracer of the r--process.  As previously mentioned, nearly all 
globular clusters are r--process rich with [Eu/Ba,La]$\approx$+0.25 (e.g., 
Gratton et al. 2004), but previous studies have shown that $\omega$ Cen has
very strong s--process enhancement, especially at [Fe/H]$\ga$--1.5 (e.g., 
Francois et al. 1988; Norris \& Da Costa 1995; Smith et al. 1995, 2000).  We
find in agreement with previous studies that most $\omega$ Cen stars more 
metal--rich than [Fe/H]$\sim$--1.7 are strongly s--process enriched based on 
[La/Eu] ratios approaching and exceeding +0.8.  While the most metal--poor 
stars have $\langle$[La/Eu]$\rangle$=--0.02, this value rises to 
$\langle$[La/Eu]$\rangle$=+0.49 by [Fe/H]$\sim$--1.4 meaning that [La/Eu] 
increases by more than a factor of 3 during a span in which [Fe/H] increases 
by a factor of 2.  However, we find that $\langle$[La/Fe]$\rangle$ does not 
increase appreciably at metallicities exceeding [Fe/H]$\sim$--1.2 (excluding
possible Ba--stars), but all stars in our sample with [Fe/H]$>$--1.2 have 
[La/Fe]$>$+0.5 and abundance patterns dominated by the s--process.  This 
trend is not shared by Eu, which remains mostly constant at 
$\langle$[Eu/Fe]$\rangle$=+0.19 ($\sigma$=0.23) at all metallicities.  We have 
also found that 25$\%$ (15/60) of our sample may qualify as Ba--stars with 
[La/Fe]$\geq$+1.0.  The most extreme case is star LEID 45358, which has 
[La/Fe]=+2.03 and [La/Eu]=+1.81.  For stars in common between the two samples, 
van Loon et al. (2007) found these to have large Ba4554 indices indicating 
they are Ba--rich as well.

\section{DISCUSSION}

\subsection{Chemical Enrichment in $\omega$ Cen}

Spectroscopic and photometric analyses of $\omega$ Cen stars have revealed a 
system hosting a complex past.  The cluster metallicity apparently increased 
from [Fe/H]$\approx$--2.2 to [Fe/H]$\approx$--0.5 over roughly 2--4 Gyrs (e.g., 
Stanford et al. 2006) and $\omega$ Cen has experienced a handful of discrete 
star formation events.  The metallicity distribution peaks in our data agree
with those found in the literature and correspond to the ``MP" 
([Fe/H]$\sim$--1.7), ``MINT2" ([Fe/H]$\sim$--1.4), ``MINT3" 
([Fe/H]$\sim$--1.0), and ``SGB--a" ([Fe/H]$\sim$--0.6) populations found on 
the SGB by Sollima et al. (2005b).  However, these classifications are not as
well--defined on the main sequence and show considerable complexity (e.g.,
Bedin et al. 2004; Piotto et al. 2005; Villanova et al. 2007).  The apparent 
kinematic homogeneity of the various stellar populations (e.g. Pancino et al.
2007; Bellini et al. 2009) suggests most, if not all, of the cluster's 
chemical enrichment is the result of internal processes rather than a product 
of multiple merger events.  However, the paucity of stars more metal--poor 
than [Fe/H]$\sim$--2 means the nascent gas from which the primary population 
formed was already considerably polluted by massive star ejecta.  One of the 
most striking results discovered so far is that the second most metal--poor 
stellar population ([Fe/H]$\sim$--1.4; and perhaps the subsequent more 
metal--rich stars) may have experienced both a huge increase in He content 
(dY/dZ$\sim$70; Piotto et al. 2005) and an equally impressive increase in 
s--process element abundances compared to the primary population 
([Fe/H]$\sim$--1.7), which contains more than half of all $\omega$ Cen stars.  
Somehow these events took place while preserving the various light element 
correlations observed in other globular clusters that do not (in general) have 
large He and metallicity variations and lack strong s--process signatures.  
Since the combined Johnson et al. (2008) and current data sets allow us to 
probe various production sources, we turn now to what our current data add to 
$\omega$ Cen's puzzling past.

\subsubsection{Supernova Pollution}

The majority of elements heavier than Li are produced during various quiescent 
and explosive nucleosynthetic events in $\ga$11 M$_{\sun}$ stars (Woosley \&
Weaver 1995).  These processes, which occur within $\la$20$\times$10$^{\rm 6}$ 
years after the onset of star formation, are known to produce an overabundance 
of $\alpha$ elements by about a factor of two relative to the solar $\alpha$/Fe
ratio.  In addition, massive stars also produce varying amounts of the odd--Z
light elements (e.g., C through Al) with metallicity dependent yields of
--0.5$\la$[X/Fe]$\la$+0.3 in the metallicity regime covered by $\omega$ Cen
stars (e.g., Woosley \& Weaver 1995; Chieffi \& Limongi 2004; Nomoto et al. 
2006).  Although the final abundances of Fe--peak elements are 
dependent on the explosion energy and mass--cut, they generally track closely 
to Fe.  These stars inevitably produce some neutron capture elements as well, 
but only the lower mass SNe ($\sim$8--11 M$_{\sun}$) are believed to be 
significant r--process contributors (e.g., Mathews \& Cowan 1990; Cowan et al. 
1991; Wheeler et al. 1998), while low mass AGB stars ($\sim$1--4 M$_{\sun}$) 
seem the best candidates for s--process production (e.g., Busso et al. 1999).

In contrast, mass transfer Type Ia SNe may take anywhere from 
500$\times$10$^{\rm 6}$ to more than 3$\times$10$^{\rm 9}$ years to evolve 
(e.g., Yoshii et al. 1996) and could have difficulty forming in low metallicity 
([Fe/H]$<$--1) environments (Kobayashi et al. 1998).  Nucleosynthesis 
calculations have shown that these SNe predominantly produce Fe--peak elements 
and only trace amounts of $\alpha$ and light elements (Nomoto et al. 1997).  It
is estimated that Type Ia SNe have produced at least 50$\%$ of the total 
\iso{56}{Fe} in the Galaxy and their onset is believed to be the primary cause 
for the decline in [$\alpha$/Fe] at [Fe/H]$>$--1 in the disk and halo 
populations.  It is for this reason that the [$\alpha$/Fe] ratio is often used 
as a diagnostic to test the presence of Type Ia SNe in a stellar system.

While there have been some $\alpha$--poor stars found in $\omega$ Cen's most 
metal--rich population (Pancino et al. 2002), the general trend of enhancement 
in the $\alpha$ elements suggests a majority of the cluster's chemical 
evolution occurred before Type Ia SNe had time to evolve.  Determining whether 
or not Type Ia SNe can form in metal--poor environments could help place 
additional constraints on $\omega$ Cen's evolutionary timescale.  If the lower 
limit of [Fe/H]$\sim$--1 estimated by Kobayashi et al. (1998) is correct and 
only the most metal--rich population in the cluster is affected by Type Ia SNe
ejecta, then this would imply an age difference between the [Fe/H]$\sim$--1
and [Fe/H]$\sim$--0.7 groups of $\la$1 Gyr.  However, if this limit is at a 
much lower metallicity, then the cluster would have had to evolve on a much
shorter time scale.

\begin{figure}
\epsscale{0.90}
\plotone{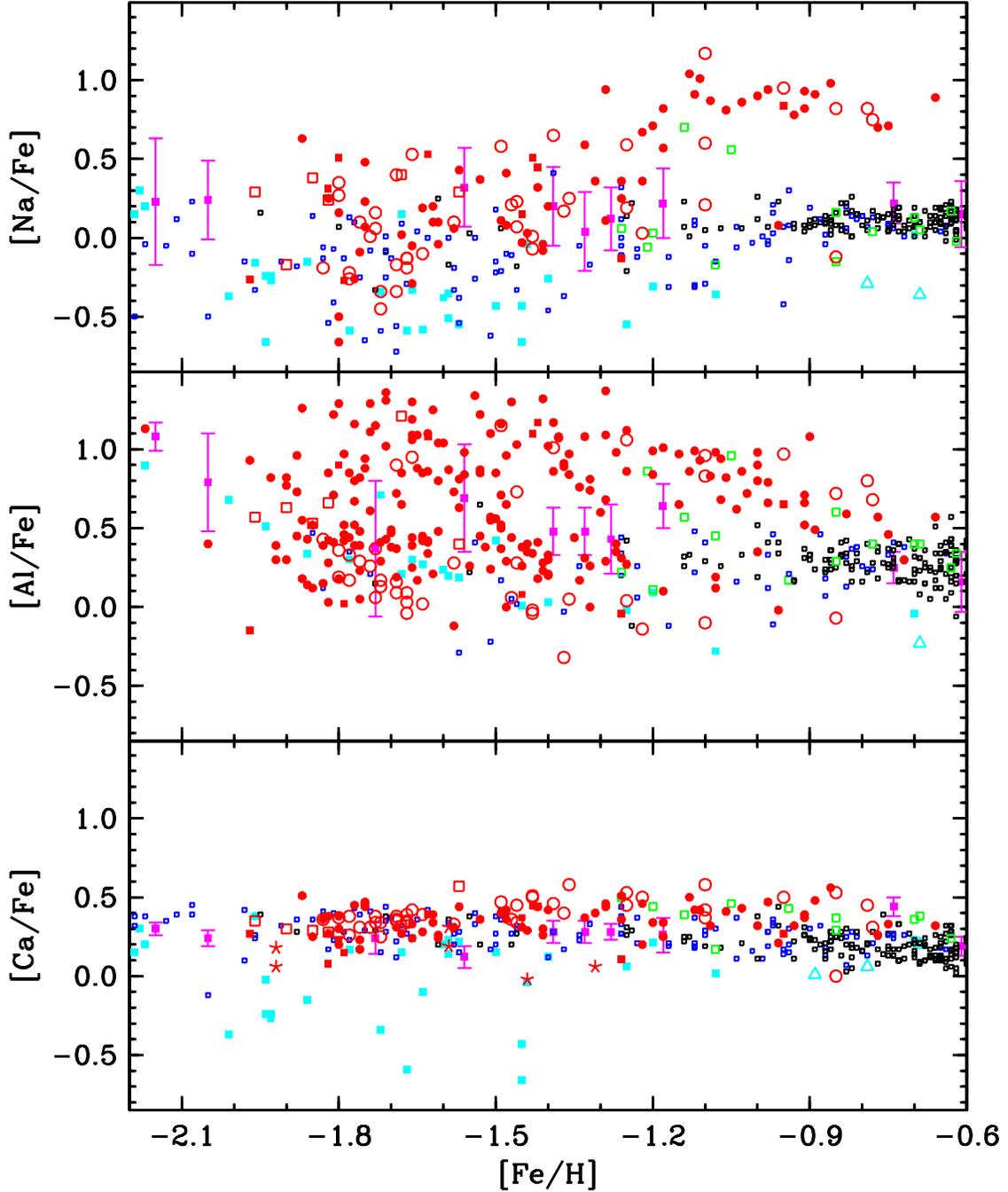}
\caption{Plots of [Na/Fe], [Al/Fe], and [Ca/Fe] versus [Fe/H] are shown with
data from this study and the literature.  The filled circles are values from
the combined sample of this study and Johnson et al. (2008), the open circles
are from Norris \& Da Costa (1995), the filled squares are from Smith et al. 
(2000), the open squares are from Francois et al. (1988), and the stars are 
from Smith et al. (1995).  Literature values are provided for the thin/thick
disk (open black boxes), halo (open blue boxes), bulge (open green boxes), 
dwarf spheroidals (filled cyan boxes), globular clusters with 1$\sigma$ bars 
(filled magenta boxes), and the Sagittarius dwarf spheroidal (open cyan 
triangles).  References are given in Table 5.}
\label{f11}
\end{figure}

\begin{figure}
\epsscale{1.00}
\plotone{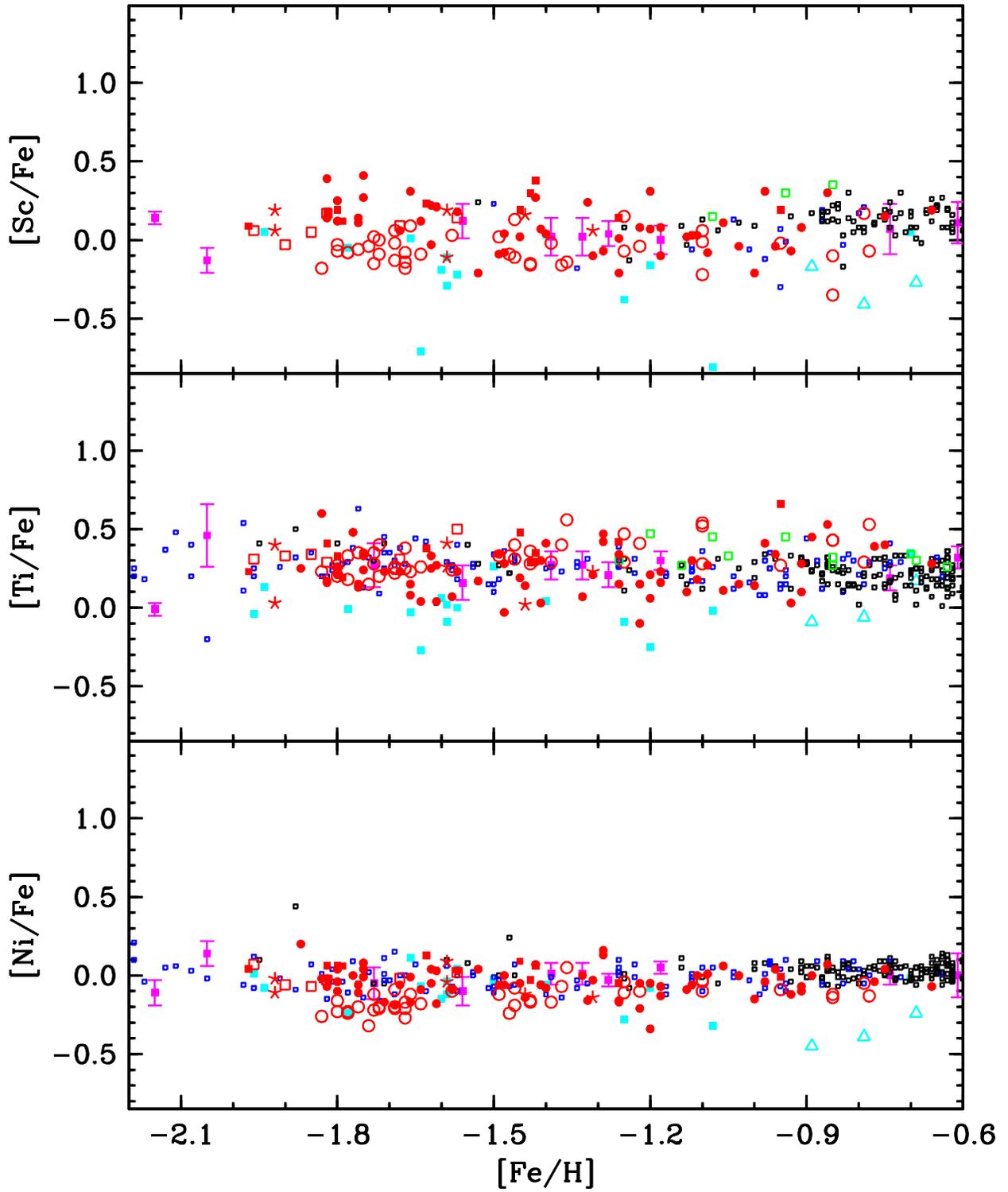}
\caption{Plots of [Sc/Fe], [Ti/Fe], and [Ni/Fe] versus [Fe/H] are shown with
data from this study and the literature.  The symbols and [X/Fe] scales are 
the same as in Figure \ref{f11}.}
\label{f12}
\end{figure}

\begin{figure}
\epsscale{1.00}
\plotone{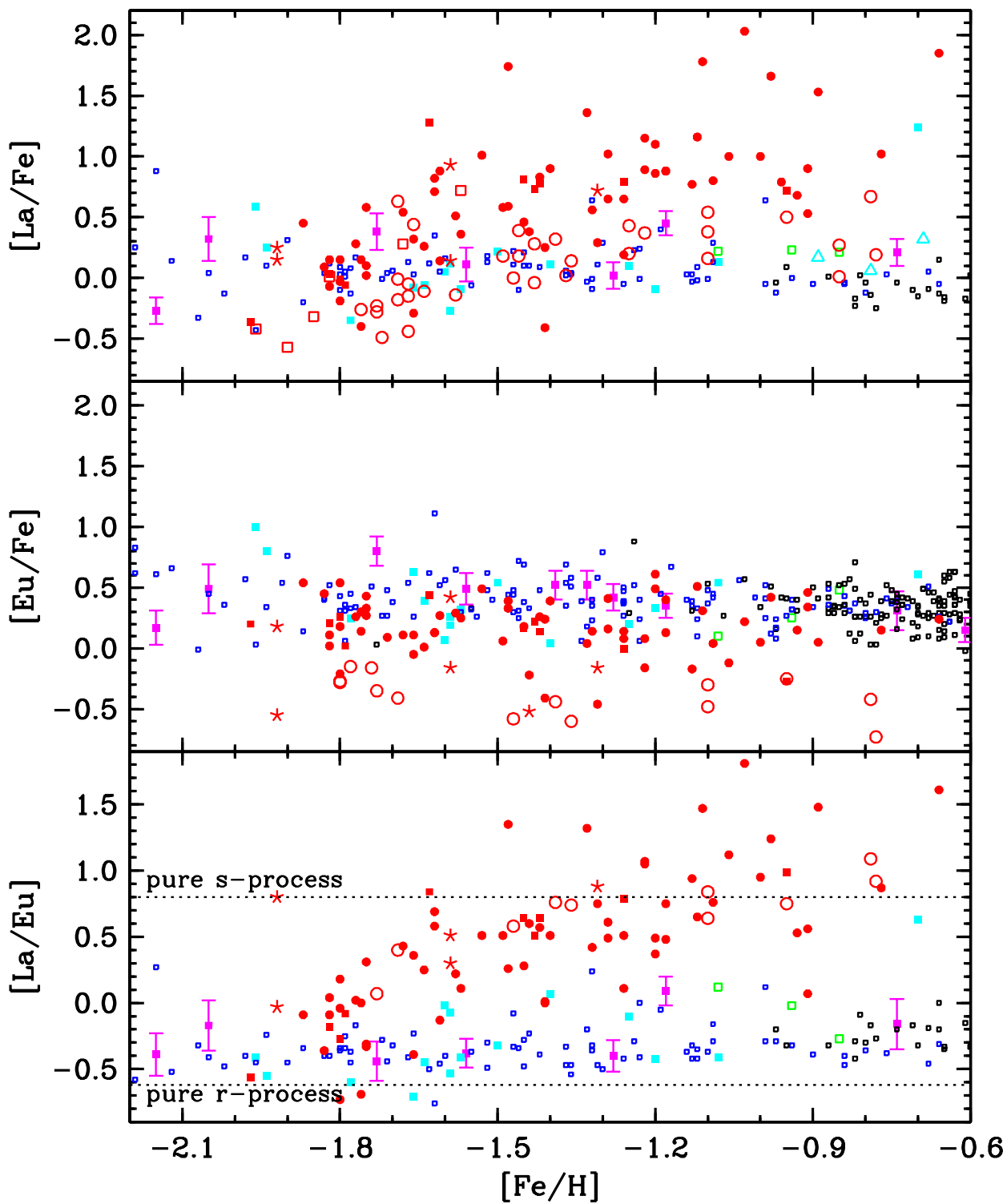}
\caption{Plots of [La/Fe], [Eu/Fe], and [La/Eu] versus [Fe/H] are shown with
data from this study and the literature.  The symbols are the same as in
Figure \ref{f11}.  The dashed lines indicating pure s--process and r--process
abundance ratios are taken from McWilliam (1997).}
\label{f13}
\end{figure}

In Figures \ref{f11}--\ref{f13} we show the evolution of all elements measured
in this study as a function of [Fe/H] along with those available in the
literature for $\omega$ Cen, the Galactic disk, bulge, halo, globular clusters,
and nearby dwarf spheroidal galaxies (see Table 5 for data references).  From 
these data we have confirmed: (1) a more than 1.0 dex spread exists for [Na/Fe]
and [Al/Fe] and the two elements are correlated, (2) the $\alpha$ elements are 
enhanced by about a factor of two at all metallicities with small star--to--star
scatter, (3) there are at least four different metallicity peaks (see also 
Figure \ref{f9}) at [Fe/H]=--1.75, --1.45, --1.05, --0.75 with internal 
dispersions of $\sim$0.10 dex in each subpopulation, and (4) there is a large 
s--process component that manifests itself in the intermediate and metal--rich 
populations of the cluster.  As is the case for other globular clusters, the 
larger star--to--star variations seen in the light and neutron--capture 
elements versus the $\alpha$ and Fe--peak elements suggest additional 
production (or destruction) sources other than core collapse SNe.  We know 
that, at least for the light elements, the observed inhomogeneity is not due 
to incomplete mixing of SN ejecta because the Na/Al enhanced stars are also 
O--poor (e.g., Norris \& Da Costa 1995; Smith et al. 2000).

If massive stars cannot account for all of the observed abundance anomalies in
$\omega$ Cen, then how much can they account for?  At least in stars with 
[Fe/H]$<$--1, Type II SNe are responsible for producing nearly all of the 
$\alpha$ and Fe--peak elements.  However, IMF weighted theoretical yields 
of SNe with initial metallicities in the range --2$<$[Fe/H]$<$--0.5 (e.g., 
Woosley \& Weaver 1995; Chieffi \& Limongi 2004; Nomoto et al. 2006) produce 
values roughly consistent with those observed in the disk, halo, and bulge 
(i.e., $\langle$[Na/Fe]$\rangle$$\sim$0; $\langle$[Al/Fe]$\rangle$$\sim$+0.3), 
but $\omega$ Cen (and other globular cluster) stars can reach [Na/Fe]$\sim$+1.0
and [Al/Fe]$\sim$+1.4.  Using the Al data in Figure \ref{f11} to trace the 
percentage of stars with light element abundance patterns matching those 
observed in the other Galactic populations at comparable metallicity 
([Al/Fe]$\leq$+0.5), we find 42$\%$ (84/202) of our sample fall into this 
category.  It is more difficult to quantify this with the Na data because the 
sample size is more than a factor of three smaller, but it appears that at 
least a significant fraction of the stars in Figure \ref{f11} with [
Fe/H]$<$--1.2 show [Na/Fe] ratios consistent with the disk, halo, and bulge, 
but nearly all of the more metal--rich stars are enhanced in Na.  This further 
solidifies the claim that although Type II SNe have had a significant impact 
on all $\omega$ Cen stars, they are not the only significant nucleosynthesis 
site.  Assuming our data are representative, roughly half of all $\omega$ Cen 
stars appear to have formed in an environment that was polluted with the 
ejecta from sources other than Type II SNe.

Further inspection of Figure \ref{f11} reveals an interesting trend in Na and 
Al as a function of [Fe/H].  As noted in $\S$4.2, there is a clear lack of 
stars showing Na and Al abundances consistent with being polluted solely by 
Type II SNe at [Fe/H]$\ga$--1.2.  Only 6$\%$ (1/17) of $\omega$ Cen giants are 
``Na--normal" ([Na/Fe]$\sim$0), and this trend is present in both the 
Norris \& Da Costa (1995) and Smith et al. (2000) data as well.  A similar 
result is observed in the larger sample of Al data in that only 22$\%$ (8/36) 
are ``Al--normal" ([Al/Fe]$\la$+0.3).  While there appears to be a down turn 
in the maximum [Al/Fe] attained at [Fe/H]$>$--1.2, the rise in [Na/Fe] and 
[La/Fe] coupled with the stability of [$\alpha$/Fe] and [Eu/Fe] in the same 
metallicity range indicates this artifact is not the result of Type Ia SNe 
adding Fe but instead a decrease in the [Al/Fe] ratio being added to the 
cluster's ISM by the production source.  What is perhaps most intriguing is 
that despite a (possible) huge increase in He between the [Fe/H]=--1.7 and 
--1.4 groups, the light element trends are very similar.  It would seem that 
whichever stars are the source of the high He abundances do not produce 
abnormally large [Na/Fe] and [Al/Fe] ratios because similar enhancements in Na 
and Al are found in globular clusters that do not show signs of such extreme 
He variations. 

If Eu production can be attributed mostly to 8--10 M$_{\sun}$ stars, then 
we know from those data alone that chemical enrichment had to have occurred in
$\omega$ Cen over more than $\sim$200$\times$10$^{\rm 6}$ years because there
are at least four subpopulations with different [Fe/H] and [Eu/Fe] is roughly 
constant (with some scatter).  However, [Eu/Fe] is, on average, consistently 
at least 0.1--0.2 dex underabundant relative to the other populations shown in
Figure \ref{f13}.  The reason for this is not clear, but it could be that the
ratio of 8--10 M$_{\sun}$ versus higher mass stars was anomalously low in 
$\omega$ Cen relative to other systems.  

\subsubsection{Intermediate Mass AGB Stars}

The discovery of significant star--to--star scatter in light elements coupled 
with the O--Na anticorrelation in stars on the main sequence and 
subgiant branches of globular clusters seems to indicate that the various 
relations among the elements O through Al were already imprinted on the gas
from which the current generations of stars formed.  As discussed in $\S$1, 
HBB occurring in intermediate mass ($\sim$5--8 M$_{\sun}$) AGB stars is 
currently favored as a likely location for producing the light element 
trends.  These stars have the advantages of preserving their initial [Fe/H]
envelope abundances, ejecting enriched material at low velocities, experiencing
few third dredge--up episodes (negligible s--process production), and reaching
envelope temperatures $>$70$\times$10$^{\rm 6}$ K that activate the NeNa 
and MgAl proton--capture cycles.  However, current AGB stellar models are 
highly sensitive to the adopted treatment of convection and mass loss and it 
has been pointed out that these scenarios do not explain the role of super--AGB
stars (those that ignite core carbon but not neon burning) nor 1--4 M$_{\sun}$
AGB stars (Prantzos \& Charbonnel 2006).  Models using standard mixing length 
theory (e.g., Fenner et al. 2004; Karakas \& Lattanzio 2007) are unable to 
reproduce the large O depletions ([O/Fe]$<$--0.6) found in some globular 
cluster stars (including $\omega$ Cen) and show large enhancements in 
[C+N+O/Fe], which conflict with observations that the CNO sum is constant 
(Pilachowski 1988; Dickens et al. 1991; Norris \& Da Costa 1995; Smith et al. 
1996; Ivans et al. 1999).  On the other hand, models adopting the full 
spectrum of turbulence treatment of convection (e.g., Ventura \& D'Antona
2008) show fewer third dredge--up episodes and thus keep the CNO sum roughly 
constant while explaining some of the C through Al abundance trends seen in 
globular clusters.  Neither case is able to fully explain all light element 
anomalies, in particular the super O--poor stars and Mg isotopic ratios, which 
may require a hybrid scenario that includes \emph{in situ} deep mixing 
and HBB in $\ga$5 M$_{\sun}$ AGB stars (e.g., D'Antona \& Ventura 2007) in
addition to improvements in key nuclear reaction rates. 

Can these stars reproduce what we observe in $\omega$ Cen?  Our data have shown
that only about half of the stars in our sample are consistent with being 
formed from gas predominantly polluted by Type II SNe (i.e., the stars are not
particularly enhanced in Na and Al compared to disk and halo stars of 
comparable metallicity).  Since $>$5 M$_{\sun}$ AGB stars likely do not alter 
the abundances of any elements heavier than Al, we will restrict the 
discussion to those elements.  First turning to the populations with 
[Fe/H]$<$--1.2, the stars with [Na/Fe]$>$0 and [Al/Fe]$>$+0.5 have 
envelope material that was likely exposed to high temperature proton--capture 
processing in an external environment.  Although the light element yields are 
sensitive to both model parameters and nuclear reaction rates, the 
Ventura \& D'Antona (2008) results indicate intermediate metallicity 
5--6.5 M$_{\sun}$ AGB stars can produce +0.30$<$[Na/Fe]$<$+0.60 and 
[Al/Fe]$\sim$+1.0, while the Fenner et al. (2004) data predict somewhat higher 
Na and lower Al abundances.  These values are consistent with the 
``intermediate" Al population that has $\langle$[Al/Fe]$\rangle$=+0.82 
($\sigma$=0.10) and suggest intermediate mass AGB stars could be responsible 
for the enhancements seen in these stars.  However, about 20$\%$ of the stars 
with [Fe/H]$<$--1.2 have [Al/Fe]$>$+1 and [Na/Fe]$>$+0.5.  These stars are not 
accounted for by current AGB models and may have undergone additional \emph{in 
situ} deep mixing or require pollution from another unknown source.

\begin{figure}
\epsscale{1.00}
\plotone{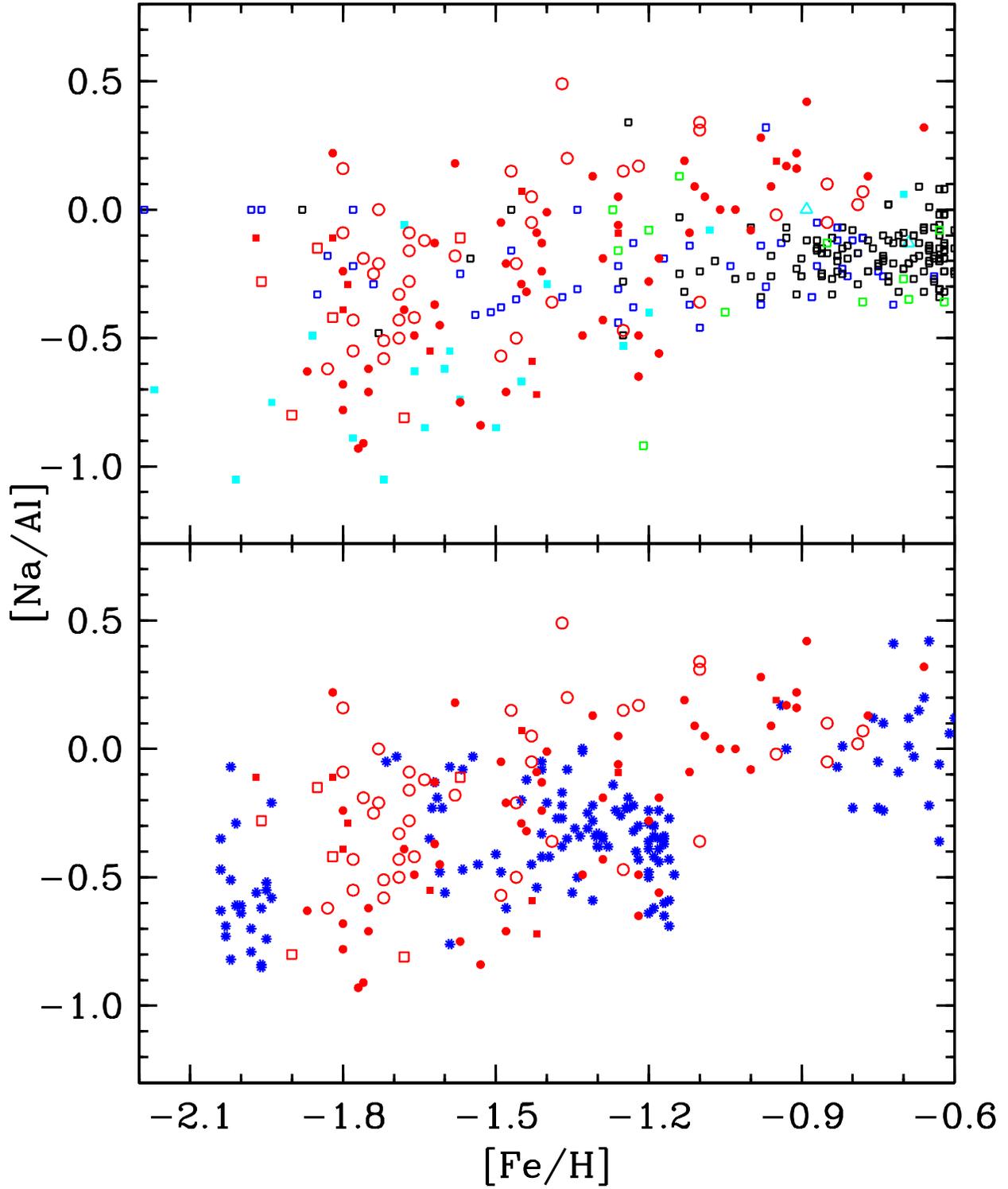}
\caption{[Na/Al] ratios as a function of metallicity are shown for a variety
of populations.  The symbols in the top panel are the same as those in
Figure \ref{f11} and the blue points in the bottom panel represent individual
globular cluster stars.}
\label{f14}
\end{figure}

As can be seen in the top panel of Figure \ref{f14}, halo, disk, and bulge 
stars exhibit a roughly constant [Na/Al]$\approx$--0.2 from [Fe/H]=--2 to 
--0.6, while $\omega$ Cen, dwarf spheroidal, and globular cluster stars 
display a wide range from [Na/Al]=--1 to +0.4 and show a general increase in 
$\langle$[Na/Al]$\rangle$ with increasing metallicity.  Since the final 
abundances of Na and Al in SN ejecta scale similarly with neutron excess and 
metallicity (Arnett 1971), the Na/Al ratio is mostly insensitive to 
metallicity changes and is consistently near [Na/Al]$\sim$--0.2 (e.g., 
Woosley \& Weaver 1995).  The overproduction of Al at low metallicities and 
underproduction at higher metallicities is consistent with the observed trends 
in AGB models (e.g., Ventura \& D'Antona 2008) due to lower temperatures 
at the bottom of the convective envelope and shallower mixing in more 
metal--rich stars.  This trend is nearly identically reproduced in globular
cluster stars of varying metallicity (bottom panel of Figure \ref{f14}) and 
likely indicates the same stars that are responsible for the globular cluster 
light element anomalies are also prevalent in $\omega$ Cen.  Since the same
trend is also observed in dwarf spheroidal stars, which are not believed to 
be strongly enriched in Type II SN ejecta, this may strengthen the case for 
HBB in intermediate mass AGB stars (or some equivalent H--burning environment) 
to be the source of light element abundance trends different than those seen 
in the disk and halo.  It is interesting that these two systems share the 
rise in [Na/Al] versus [Fe/H] with globular clusters because, as their low 
[$\alpha$/Fe] ratios indicate, star formation has proceeded much differently 
in dwarf spheroidals despite having comparable main sequence turnoff age ranges 
with $\omega$ Cen.  

\begin{figure}
\epsscale{1.00}
\plotone{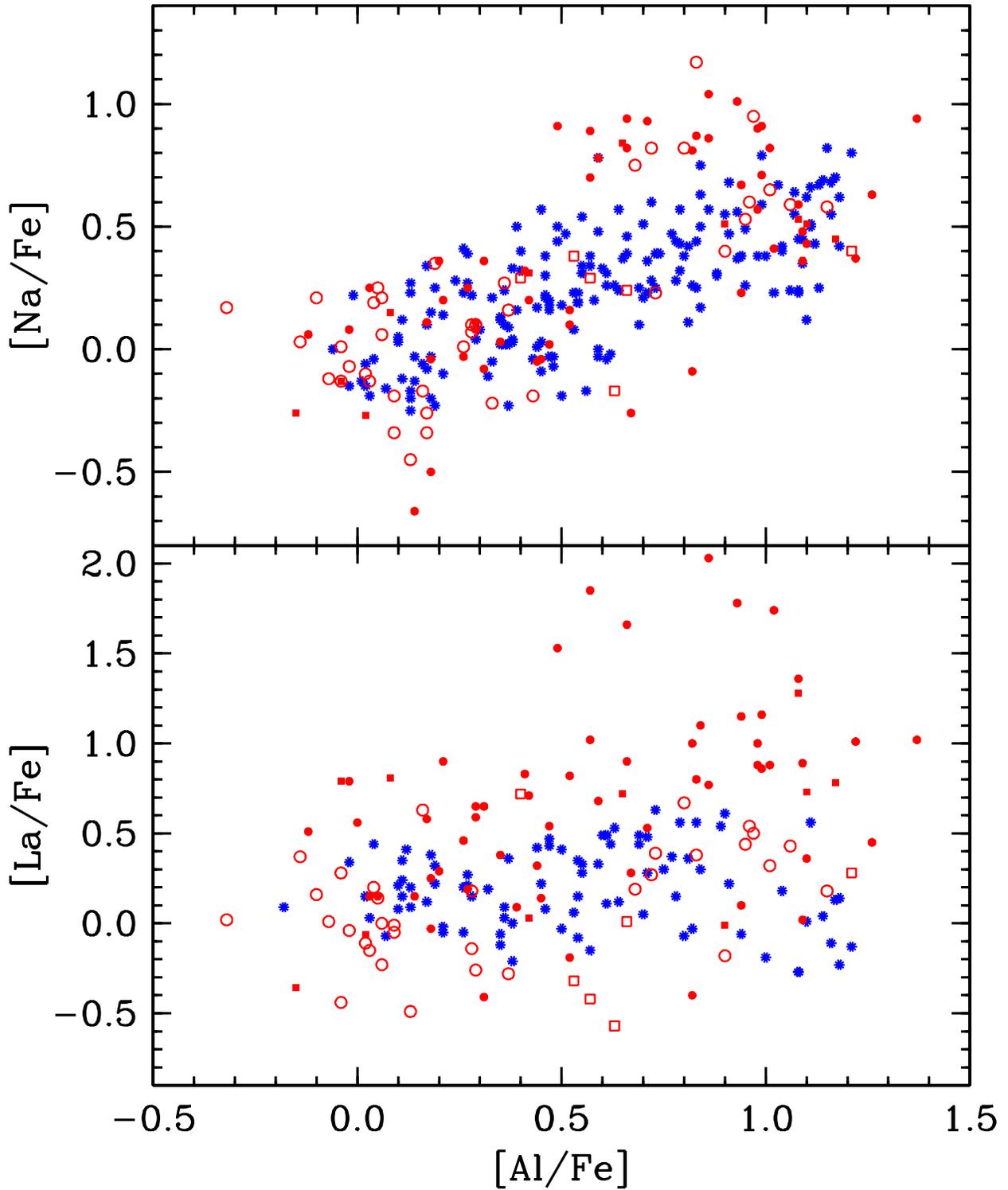}
\caption{The top panel shows [Na/Fe] versus [Al/Fe] and compares $\omega$ Cen
data to results from individual globular cluster stars.  The bottom panel
shows the same set of stars but plots [La/Fe] versus [Al/Fe].  The symbols
are the same as those in Figure \ref{f14}.}
\label{f15}
\end{figure}

The paucity of stars with [Al/Fe]$>$+1 at [Fe/H]$>$--1.2 is also consistent 
with the predictions of \emph{in situ} deep mixing at higher metallicities
where the increased $\mu$--gradient is expected to inhibit dredge--up of 
ON, NeNa, and MgAl cycled material into the stellar envelope via meridional
circulation (e.g., Sweigart \& Mengel 1979).  While the range in Na and Al 
data track closely to that of other globular clusters at low and intermediate
metallicity, the more metal--rich $\omega$ Cen stars show surprising Na 
enhancements and decreased star--to--star scatter that are not seen in 
globular clusters of comparable metallicity.  This is true even for M4 
([Fe/H]$\sim$--1.1), which is suspected of having a second, more enriched 
population without a large spread in Fe (Marino et al. 2008).  Although the 
range of M4's Al abundances are consistent with the values we find here, the 
\emph{average} [Na/Fe] ratio in $\omega$ Cen giants of comparable metallicity 
is about 0.3 dex larger than the highest [Na/Fe] abundance found by either 
Ivans et al. (1999) or Marino et al. (2008) in M4.  It would seem that there 
was an additional source of Na in the more metal--rich $\omega$ Cen 
populations or that hardly any unenriched gas remained to dilute the AGB 
ejecta.  Figure \ref{f15} illustrates this point in that the stars with
[Fe/H]$>$--1.2 and [Al/Fe]$>$+0.5 have [Na/Fe] ratios that lie above the range
expected for a given Al abundance based on typical globular cluster values.
The identity of the Na source is only speculative, but if the 
progenitor AGB population that polluted the gas from which the [Fe/H]$>$--1.2 
stars formed was He--rich, the higher temperatures and possible deeper 
mixing in regions where the NeNa cycle was operating may have contributed to 
the increased Na abundances.  It may also be possible that lower mass, He--rich
AGB stars, which evolve more quickly than He--normal stars (and produce more Na
and less Al), could have a larger impact than in normal globular clusters.
However, $\la$4 M$_{\sun}$ AGB stars are not believed to strongly deplete O 
and would have to already be O--poor to reproduce the sub--solar [O/Fe] ratios
found in many $\omega$ Cen giants with [Fe/H]$\ga$--1.5.

\subsubsection{Low Mass AGB Stars}

Lower mass, thermally pulsing AGB stars ($\sim$1--4 M$_{\sun}$), which evolve 
over 150$\times$10$^{\rm 6}$ to 2.5$\times$10$^{\rm 9}$years (Schaller et al. 
1992), are thought to be the primary producers of the main s--process 
component in the Galaxy at metallicities found in $\omega$ Cen (e.g., Busso et 
al. 2001).  Smith et al. (2000) showed that the [Rb/Zr] ratio in $\omega$ Cen 
was consistent with the s--process being produced in 1.5--3.0 M$_{\sun}$ AGB 
stars, implying a monotonic, total evolutionary timescale of $\sim$2--3 Gyrs.  
This is consistent with most other estimates (e.g., Stanford et al. 2006; but 
see also Villanova et al. 2007).  Since these stars have the longest formation 
timescale, the presence of their chemical signatures sets a lower limit on 
relative age estimates.

The halo and disk populations are known to exhibit a steady rise in the 
contribution of s--process elements at [Fe/H]$>$--2.5 (e.g., Simmerer et al. 
2004), but globular cluster heavy element abundances are dominated by the 
r--process (e.g., Gratton et al. 2004) and are indicative of the rather rapid 
chemical evolution timescales of normal globular clusters compared to the disk 
and halo.  Interestingly, dwarf spheroidal stars tend to have a stronger 
s--process component than any of the Galactic populations (e.g., Geisler et 
al. 2007), but one that is much smaller than that seen in $\omega$ Cen.  This,
along with the evidence for Type Ia SN pollution, implies dwarf spheroidal 
galaxies evolve much differently than most other Galactic stellar systems and 
do so with a rather subdued star formation rate (e.g., Mateo 2008).  

However, the Galactic bulge is believed to have formed rapidly, as constrained
by turnoff photometry (Ortolani et al. 1995; Kuijken \& Rich 2002; Zoccali et 
al. 2003; Clarkson et al. 2008) as well as by measured high [$\alpha$/Fe] 
(e.g., McWilliam \& Rich 1994; Fulbright et al. 2006; Lecureur et al. 2007).
Theoretical studies argue for timescales significantly less than 
10$^{\rm 9}$ yrs (e.g., Elmegreen 1999; 2008; Ballero et al. 2007).  Yet 
despite a metallicity that is high compared to the halo and $\omega$ Cen, 
(e.g., Fulbright et al. 2006, Zoccali et al. 2008) the s--process elements are 
seen to exhibit Solar [X/Fe] ratios (McWilliam \& Rich 1994) that would appear 
to require low and intermediate mass stars to have provided significant input 
to the bulge's chemical evolution.

In Figure \ref{f13} we show the evolution of [La/Fe], [Eu/Fe], and [La/Eu] as
a function of [Fe/H] for $\omega$ Cen and other stellar populations.  Since 
all but the most metal--poor group of $\omega$ Cen stars show significant 
enhancement in the s--process element La (and the [La/Eu] ratio), we find in 
agreement with previous studies that at least 10$^{\rm 9}$ years had to have 
passed between the formation of the primary population at [Fe/H]=--1.7 and the 
final population at [Fe/H]=--0.7 to allow the low mass progenitor populations 
enough time to evolve.  A significant percentage (25$\%$) of stars in our 
sample have [La/Fe]$\geq$+1.0 and may be the result of binary mass transfer 
from a $<$4 M$_{\sun}$ AGB companion.  However, none of these stars are 
present in the dominant, most metal--poor population but are found at 
[Fe/H]$>$--1.5 with most being present at [Fe/H]$>$--1.  It is unknown whether 
the prevalence of such stars at higher metallicities is a result of the longer 
formation timescales needed for one of the companions to evolve, an anomalous 
increase in the binary fraction at higher metallicity, or a sample selection 
effect.  If the result is not a selection effect, then this may be a clear 
indication that the more metal--rich stars are at least 
1--2$\times$10$^{\rm 9}$ years older than the metal--poor population. 

\begin{figure}
\epsscale{1.00}
\plotone{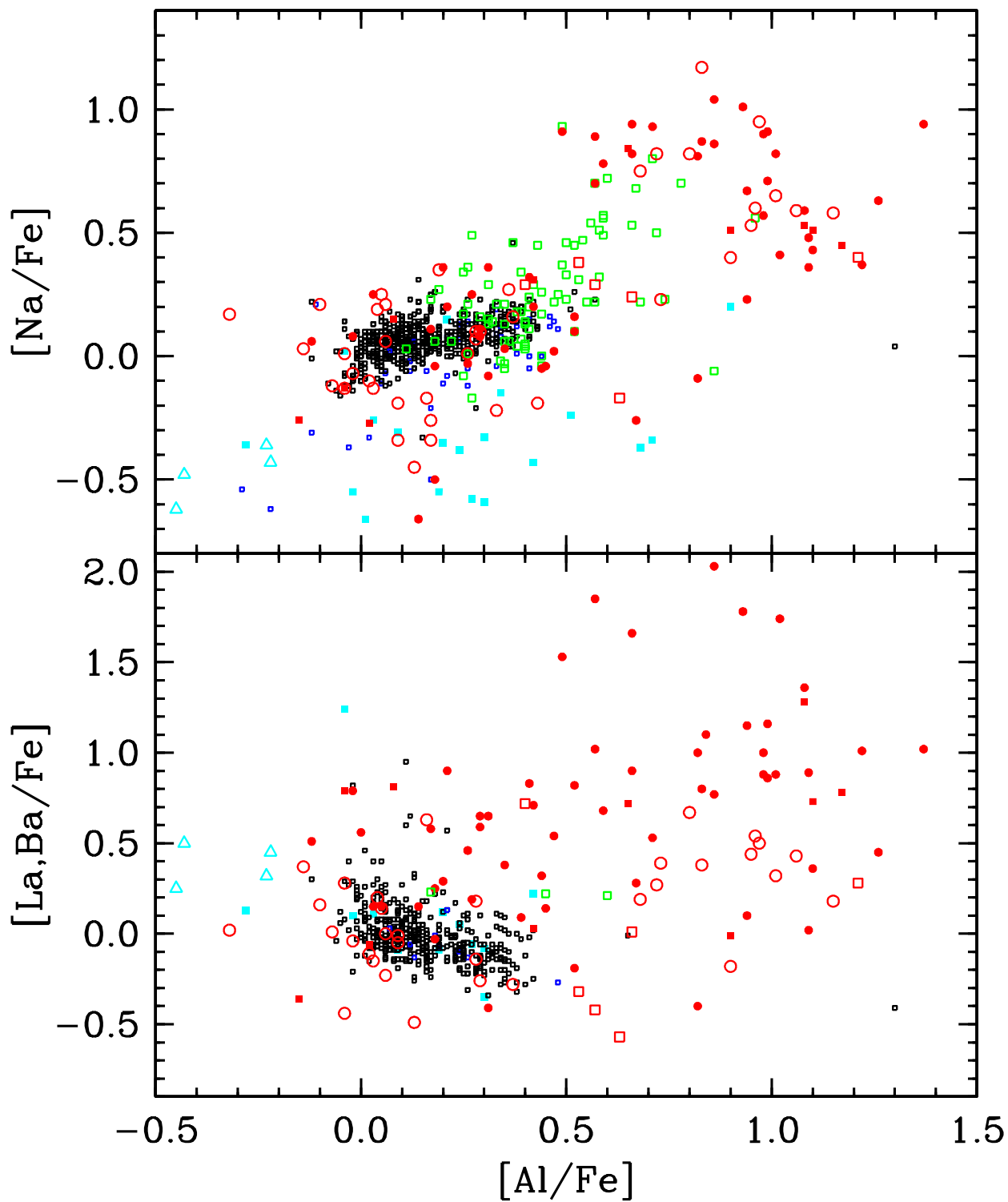}
\caption{The top panel shows [Na/Fe] versus [Al/Fe] with data from this study
and the literature.  The bottom panel shows [La,Ba/Fe] versus [Al/Fe] where 
the Ba data are used as a tracer for the s--process in the disk and halo while
La is used in all other cases.  The symbols are the same as those in Figure 
\ref{f11}.}
\label{f16}
\end{figure}

Figures \ref{f15} and \ref{f16} show [Na/Fe] and [La/Fe] versus [Al/Fe], which
could be a useful indicator regarding the relative importance of low versus 
intermediate mass AGB stars.  In most globular clusters there is little 
evidence of light elements showing any correlation with heavy neutron--capture 
elements on top of the correlations seen among the various light elements 
(e.g., Smith 2008), which implies the elements lighter than Al are produced in 
a different astrophysical site over different timescales than those produced 
via the s--process and r--process.  This may mean that the current generation 
of globular cluster stars have abundance signatures strongly weighted towards 
pollution from more massive AGB stars compared to those $\la$4 M$_{\sun}$.  On 
the contrary, $\omega$ Cen exhibits a mild correlation between La and Al (as 
well as Na) and as stated above shows [La/Fe] ratios well in excess of the 
roughly [La/Fe]$\sim$+0.5 maximum found in globular cluster stars, especially 
at [Fe/H]$>$--1.5.  Current AGB nucleosynthesis models (e.g., Ventura \& 
D'Antona 2008) suggest that this correlation is unlikely to be the result of 
$\sim$1--4 M$_{\sun}$ stars dominating the chemical enrichment of $\omega$ Cen 
because AGB stars in that mass range are shown to produce Na without 
significantly depleting O, which contradicts the O--Na anticorrelation observed
in the cluster giants and prevalence of O--poor stars at higher metallicities 
(e.g., Norris \& Da Costa 1995).  For the other populations shown in Figure 
\ref{f16}, only the bulge data show any hint of a Na--Al correlation, but that 
is not believed to be the result of the same mechanism at work in globular 
clusters (Lecureur et al. 2007).  However, the current lack of heavy element 
data in the bulge makes it difficult to draw conclusions regarding the impact 
of low mass AGB stars in that environment.  Since none of the other populations
show any correlation between La and Al (or Na), it appears that $\omega$ 
Cen is (as always) a special case where both low and intermediate mass AGB 
stars have had significant influence on the cluster's chemical evolution.

In this paper and previous studies, it has been shown that $\omega$ Cen is an
extremely complex object with an intriguing formation history.  Nearly all 
aspects of its past remain a mystery and although it has been shown that the
cluster experienced multiple star formation episodes (and probably significant
mass loss), there is evidence both for and against simple monotonic chemical
enrichment (i.e., metal--poor stars are older than more metal--rich stars).  It
appears that $\omega$ Cen shares many chemical characteristics with a variety
of systems that formed under widely different conditions and the cluster 
exhibits signs of both rapid and extended star formation.  One of the
interesting issues raised by our data is the significance of the apparent 
transition in light element abundance trends at [Fe/H]$\approx$--1.2.  It seems
as if the stars with [Fe/H]$>$--1.2 were made almost entirely out of AGB 
ejecta, but the populations with [Fe/H]$<$--1.2 contain groups of stars that 
likely formed both with and without the presence of AGB pollution in nearly 
equal proportions.  The lack of $\alpha$--poor stars in all but perhaps the 
most metal--rich population poses a serious problem and $\omega$ Cen's 
enrichment history challenges the paradigm of chemical evolution that for 
timescales $>$1 Gyr, Type Ia SNe contribute Fe--peak and $\alpha$--poor 
material that drive down the [$\alpha$/Fe] ratio to near solar composition.  It
may be that the cluster lost too much mass before the onset of Type Ia SNe or 
the ejecta were located too far outside the core to be retained.  This may be 
corroborated by evidence that there is no radial preference in the location of 
X--ray binaries in $\omega$ Cen due to a lack of mass segregation (e.g., Gendre
et al. 2003).  While the observation of large numbers of RGB, SGB, and main 
sequence stars are needed to understand the full picture of $\omega$ Cen's 
evolution, the large fluctuations in light element abundances such as Na and 
Mg, which are often used as metallicity tracers, make low resolution or 
integrated light studies difficult to decipher.  However, future large sample, 
high resolution studies spanning both the giant branches and main sequences 
should help further isolate the chemical signatures of each subpopulation and 
allow more quantitative analyses.

\section{SUMMARY}

We have determined abundances of several light, $\alpha$, Fe--peak, and 
neutron--capture elements for 66 RGB stars in the globular cluster $\omega$ 
Cen using moderate resolution (R$\approx$18,000) spectra.  Two different Hydra
spectrograph setups were employed spanning 6000--6250~\AA\ and 6530--6800~\AA,
yielding co--added S/N ratios of about 50--200.  The observations covered the 
full cluster metallicity regime with an emphasis on the intermediate and 
metal--rich populations.  The elemental abundances were determined using 
either equivalent width analyses or spectrum synthesis, with the addition of 
hyperfine structure data when available.  

The light elements Na and Al show large abundance inhomogeneities that span 
more than a factor of 10 and the elements are correlated.  The Al data set 
was supplemented with that from Johnson et al. (2008) and yielded [Fe/H] and 
[Al/Fe] abundances for more than 200 RGB stars.  From these data we find 
evidence for the existence of possibly three different populations of stars 
with distinct [Al/Fe] patterns.  The three sequences segment into those with 
$\langle$[Al/Fe]$\rangle$=+0.34 ($\sigma$=0.14), 
$\langle$[Al/Fe]$\rangle$=+0.82 ($\sigma$=0.10), and
$\langle$[Al/Fe]$\rangle$=+1.17 ($\sigma$=0.11) and represent 48$\%$, 34$\%$, 
and 18$\%$ of our sample, respectively.  These may be inherently tied to the 
``primordial," ``intermediate," and ``extreme" populations found in normal 
globular clusters that exhibit varying degrees of O depletion and Na 
enhancement.  However, there appears to be a break in the distribution
of both Na and Al at [Fe/H]$\approx$--1.2.  Stars with [Fe/H]$<$--1.2 have 
abundances in the range --0.1$\la$[Al/Fe]$\la$+1.4 and 
--0.5$\la$[Na/Fe]$\la$+0.6 with at least half of the stars exhibiting light 
element abundances consistent with the disk and halo populations, but more 
than 75$\%$ of stars with [Fe/H]$>$--1.2 are enhanced in Na and Al with 
values exceeding those found in the disk, halo, and even some globular 
clusters.  None of the stars with [Al/Fe]$>$+1.0 are found at [Fe/H]$>$--1.2.  
A two--sided K--S test reveals the Na and Al abundances on either side of the 
[Fe/H]=--1.2 cutoff to have a $>$90$\%$ probability of being drawn from 
different parent populations.

All of our program stars are enhanced in $\alpha$ elements with 
$\langle$[Ca/Fe]$\rangle$=+0.36 ($\sigma$=0.09) and 
$\langle$[Ti/Fe]$\rangle$=+0.23 ($\sigma$=0.14), despite showing a range of 
more than a factor of 30 in [Fe/H].  The Fe--peak elements share the same 
small range in star--to--star scatter but give roughly solar--scaled values of 
$\langle$[Sc/Fe]$\rangle$=+0.09 ($\sigma$=0.15) and 
$\langle$[Ni/Fe]$\rangle$=--0.04 ($\sigma$=0.09).  Our results are in agreement
with previous studies as we find multiple peaks in the metallicity 
distribution function at [Fe/H]=--1.75, --1.45, --1.05, and --0.75 and few 
stars with [Fe/H]$<$--1.8.  These populations represent about 55$\%$, 30$\%$, 
10$\%$, and 5$\%$ of our sample, respectively.  Additionally, we find evidence 
supporting the idea that the most metal--rich stars are more centrally 
concentrated, and there appears to be a decrease in the star--to--star 
metallicity dispersion as a function of increasing distance from the cluster 
core.

The neutron--capture elements La and Eu yield abundances indicative of strong
s--process enrichment in all but the most metal--poor stars.  We find that 
nearly all $\omega$ Cen stars with [Fe/H]$>$--1.5 have [La/Eu]$\ga$+0.5, which 
contradicts the generally r--process dominated nature of normal globular 
cluster stars that have $\langle$[La/Eu]$\rangle$$\approx$--0.25.  Despite
the sharp rise in [La/Fe], the Eu abundance remains fairly constant across all 
metallicities with [Eu/Fe]=+0.19 ($\sigma$=0.23).  However, 25$\%$ of our 
sample contains stars with [La/Fe]$\geq$+1.0 that are possibly the result of 
mass transfer in a binary system.  These stars are also known to have large 
Ba4554 indices and are predominantly found at [Fe/H]$>$--1.3.

Comparing these results with the abundance trends observed in the Galactic 
halo, disk, bulge, globular clusters, and nearby dwarf spheroidal galaxies
indicates the current generation of $\omega$ Cen stars share many chemical
characteristics found in each of those populations but contain key differences.
The elevated [$\alpha$/Fe] and solar--scaled Fe--peak abundances suggest that 
Type II SNe have dominated the production of metals in the cluster with almost 
no contribution from Type Ia SNe.  However, we find that at least 40--50$\%$ 
of stars in our sample have [Na/Fe] and [Al/Fe] ratios that exceed the yields 
expected from moderately metal--poor SNe.  Previous studies have shown that 
the Na and Al enhanced stars are also O--poor, which implies that these stars
were polluted by material that has been exposed to high temperature 
proton--capture burning.  This is corroborated by examining the behavior of 
[Na/Al] as a function of metallicity.  Type II SNe are expected to produce
a nearly metallicity independent yield of [Na/Al]$\sim$--0.2 over 
--2$<$[Fe/H]$<$--0.5, which matches observations of disk and halo stars, but 
$\omega$ Cen, normal globular cluster, and dwarf spheroidal stars span a range 
of --1$\la$[Na/Al]$\la$+0.4.  Therefore, our data strongly support the idea of 
an additional source of light elements in these environments.

HBB occurring in intermediate mass AGB stars is a favored location for 
producing Na and Al while destroying O.  Current AGB nucleosynthesis models
predict our observed trends, that more Al is produced at low metallicity and
more Na produced at high metallicity, and may explain stars with 
+0.5$<$[Al/Fe]$<$+1.0.  However, they may not be adequate to reproduce the 
$\sim$20$\%$ of metal--poor stars with [Al/Fe]$>$+1, which may require some 
other source (e.g., \emph{in situ} mixing or massive rotating stars).  What is 
perhaps most intriguing is that we find evidence for two different 
subpopulations separated as being either more metal--poor or metal--rich than 
[Fe/H]$\approx$--1.2.  Most of the stars with [Fe/H]$>$--1.2 appear to have 
formed almost entirely out of AGB ejecta and have [Na/Fe] and [Al/Fe] 
abundances well above those found in the disk and halo at similar
metallicity, while those at [Fe/H]$<$--1.2 show more of a continuum between 
strong SN pollution and AGB pollution.  Since we did not choose targets based 
on known chemical properties (e.g., CN strength), it seems that the prevalence
of Na and Al enhanced stars at higher metallicity is likely not a selection
effect.  Interestingly, although all $\omega$ Cen giants exhibit the same 
Na--Al correlation found in other globular clusters, the $\omega$ Cen stars 
with [Fe/H]$>$--1.2 have more Na for a given Al abundance by $>$0.2 dex 
compared to what is expected based on the trend seen in normal globular 
clusters.  There is also a mild correlation between La and both Na and Al, but 
it is unclear how La relates to these elements.  The decreasing maximum
value of [Al/Fe] at [Fe/H]$>$--1.2 is not shared by Na and La and suggests
a decrease in the [Al/Fe] abundance being added to the cluster's ISM rather 
than an increase in Fe due to Type Ia SNe.

The sharp increase in the abundance of [La/Fe] and [La/Eu] with increasing 
metallicity coupled with the relatively long lifetimes of stars thought to 
produce most of the s--process elements is consistent with the generally 
adopted chemical evolution timescale of $\sim$2--4 Gyr.  However, other 
stellar systems that evolved over $>$1 Gyr exhibit the characteristic downturn
in [$\alpha$/Fe], but this trend is mostly absent in $\omega$ Cen stars.  Even
though it is highly probable that $\omega$ Cen did not evolve as a closed 
box, the apparent preferential retention of Type II versus Type Ia SN ejecta or
even the suppression of Type Ia SNe at [Fe/H]$>$--1 at timescales exceeding 
1--2 Gyrs remains an important problem.

\acknowledgments

This publication makes use of data products from the Two Micron All Sky Survey,
which is a joint project of the University of Massachusetts and the Infrared 
Processing and Analysis Center/California Institute of Technology, funded by 
the National Aeronautics and Space Administration and the National Science 
Foundation. This research has made use of NASA's Astrophysics Data System 
Bibliographic Services.  RMR acknowledges support from grant AST--0709479
from the National Science Foundation.  Support of the College of Arts and 
Sciences and the Daniel Kirkwood fund at Indiana University Bloomington for 
CIJ is gratefully acknowledged.  We would like to thank the referee for his/her
thoughtful comments that led to improvement of the manuscript.

{\it Facilities:} \facility{CTIO}

\clearpage

\clearpage
\pagestyle{empty}
\setlength{\voffset}{2cm}

\tablenum{1}
\tablecolumns{15}
\tablewidth{0pt}

\begin{deluxetable}{ccccccccccccccc}
\rotate
\tablecaption{Photometry, Membership, and Model Atmosphere Parameters}
\tablehead{
\colhead{Star}	&
\colhead{Alt.}	&
\colhead{V}	&
\colhead{B$-$V}	&
\colhead{J}	&
\colhead{H}	&
\colhead{K}	&
\colhead{M$_{\rm V}^{\rm 0}$}	&
\colhead{Mem.}	&
\colhead{T$_{\rm eff}$}	&
\colhead{log g}	&
\colhead{[Fe/H]}	&
\colhead{V$_{\rm t}$}	&
\colhead{S/N}	&
\colhead{S/N}	\\
\colhead{LEID\tablenotemark{a}}	&
\colhead{ROA\tablenotemark{b}}	&
\colhead{}	&
\colhead{}	&
\colhead{}	&
\colhead{}	&
\colhead{}	&
\colhead{}	&
\colhead{Prob.\tablenotemark{c}}	&
\colhead{(K)}	&
\colhead{(cm s$^{\rm -2}$)}	&
\colhead{Avg.}	&
\colhead{(km s$^{\rm -1}$)}	&
\colhead{6125~\AA}	&
\colhead{6670~\AA}
}

\startdata

9	&	370	&	12.529	&	1.250	&	10.382	&	9.755	&	9.627	&	$-$1.543	&	99	&	4505	&	1.20	&	$-$1.41	&	2.05	&	100	&	100	\\
6017	&	240	&	12.233	&	1.420	&	9.717	&	8.982	&	8.808	&	$-$1.839	&	98	&	4145	&	0.85	&	$-$1.22	&	2.00	&	125	&	100	\\
12013	&	394	&	12.579	&	1.319	&	10.242	&	9.560	&	9.402	&	$-$1.493	&	98	&	4305	&	1.10	&	$-$1.31	&	1.85	&	100	&	125	\\
15023	&	234	&	12.182	&	1.166	&	9.964	&	9.352	&	9.231	&	$-$1.890	&	100	&	4455	&	1.05	&	$-$1.80	&	2.05	&	125	&	150	\\
16015	&	213	&	12.127	&	1.122	&	9.979	&	9.373	&	9.210	&	$-$1.945	&	100	&	4510	&	1.05	&	$-$1.80	&	1.60	&	125	&	125	\\
17015	&	325	&	12.430	&	1.156	&	10.235	&	9.610	&	9.497	&	$-$1.642	&	100	&	4470	&	1.15	&	$-$1.82	&	1.65	&	150	&	100	\\
19062	&	464	&	12.803	&	1.144	&	10.601	&	10.001	&	9.872	&	$-$1.269	&	98	&	4470	&	1.30	&	$-$1.75	&	1.50	&	100	&	100	\\
22037	&	307	&	12.339	&	1.186	&	10.178	&	9.559	&	9.402	&	$-$1.733	&	100	&	4485	&	1.10	&	$-$1.76	&	1.90	&	150	&	150	\\
\enddata

\tablenotetext{a}{Identifier from van Leeuwen et al. (2000).}
\tablenotetext{b}{Identifier from Woolley (1966).}
\tablenotetext{c}{Membership probability from van Leeuwen et al. (2000).}

\end{deluxetable}

\clearpage
\pagestyle{empty}
\setlength{\voffset}{2cm}

\tablenum{2a}
\tablecolumns{15}
\tablewidth{0pt}

\begin{deluxetable}{ccccccccccccccc}
\rotate
\tablecaption{Linelist and Equivalent Widths}
\tablehead{
\colhead{$\lambda$}	&
\colhead{Element}	&
\colhead{E.P.}	&
\colhead{log gf}	&
\colhead{9}	&
\colhead{6017}     &
\colhead{12013}     &
\colhead{15023}     &
\colhead{16015}     &
\colhead{17015}     &
\colhead{19062}     &
\colhead{22037}     &
\colhead{23061}     &
\colhead{24027}     &
\colhead{24056}     \\
\colhead{(\AA)}	&
\colhead{}	&
\colhead{(eV)}	&
\colhead{}	&
\colhead{}      &
\colhead{}      &
\colhead{}      &
\colhead{}      &
\colhead{}      &
\colhead{}      &
\colhead{}      &
\colhead{}      &
\colhead{}      &
\colhead{}      &
\colhead{}
}

\startdata

6003.01	&	Fe I	&	3.88	&	$-$1.07	&	\nodata	&	\nodata	&	\nodata	&	\nodata	&	\nodata	&	\nodata	&	\nodata	&	40	&	\nodata	&	\nodata	&	46	\\
6007.32	&	Ni I	&	1.68	&	$-$3.35	&	\nodata	&	70	&	\nodata	&	\nodata	&	\nodata	&	\nodata	&	\nodata	&	21	&	\nodata	&	\nodata	&	22	\\
6007.97	&	Fe I	&	4.65	&	$-$0.70	&	\nodata	&	53	&	\nodata	&	\nodata	&	\nodata	&	\nodata	&	\nodata	&	\nodata	&	\nodata	&	19	&	\nodata	\\
6008.57	&	Fe I	&	3.88	&	$-$0.94	&	\nodata	&	99	&	\nodata	&	\nodata	&	\nodata	&	\nodata	&	\nodata	&	\nodata	&	\nodata	&	\nodata	&	\nodata	\\
6015.25	&	Fe I	&	2.22	&	$-$4.66	&	\nodata	&	17	&	12	&	\nodata	&	\nodata	&	\nodata	&	\nodata	&	\nodata	&	\nodata	&	\nodata	&	\nodata	\\
6024.07	&	Fe I	&	4.55	&	$-$0.06	&	84	&	107	&	89	&	59	&	52	&	59	&	61	&	58	&	60	&	\nodata	&	49	\\
6027.06	&	Fe I	&	4.07	&	$-$1.14	&	47	&	68	&	63	&	25	&	15	&	24	&	35	&	22	&	28	&	57	&	20	\\
6035.35	&	Fe I	&	4.29	&	$-$2.56	&	\nodata	&	\nodata	&	\nodata	&	\nodata	&	\nodata	&	\nodata	&	\nodata	&	\nodata	&	\nodata	&	\nodata	&	\nodata	\\
\enddata

\end{deluxetable}

\clearpage
\pagestyle{empty}
\setlength{\voffset}{2cm}

\tablenum{2b}
\tablecolumns{15}
\tablewidth{0pt}

\begin{deluxetable}{ccccccccccccccc}
\rotate
\tablecaption{Linelist and Equivalent Widths}
\tablehead{
\colhead{$\lambda$}	&
\colhead{Element}	&
\colhead{E.P.}	&
\colhead{log gf}	&
\colhead{25026}	&
\colhead{26022}     &
\colhead{26086}     &
\colhead{28092}     &
\colhead{30013}     &
\colhead{32169}     &
\colhead{32171}     &
\colhead{33099}     &
\colhead{34029}     &
\colhead{34225}     &
\colhead{35172}     \\
\colhead{(\AA)}	&
\colhead{}	&
\colhead{(eV)}	&
\colhead{}	&
\colhead{}      &
\colhead{}      &
\colhead{}      &
\colhead{}      &
\colhead{}      &
\colhead{}      &
\colhead{}      &
\colhead{}      &
\colhead{}      &
\colhead{}      &
\colhead{}
}

\startdata

6003.01	&	Fe I	&	3.88	&	$-$1.07	&	\nodata	&	\nodata	&	88	&	\nodata	&	\nodata	&	\nodata	&	\nodata	&	\nodata	&	\nodata	&	\nodata	&	\nodata	\\
6007.32	&	Ni I	&	1.68	&	$-$3.35	&	18	&	\nodata	&	67	&	16	&	\nodata	&	\nodata	&	50	&	57	&	\nodata	&	\nodata	&	\nodata	\\
6007.97	&	Fe I	&	4.65	&	$-$0.70	&	\nodata	&	\nodata	&	67	&	34	&	\nodata	&	48	&	\nodata	&	\nodata	&	\nodata	&	\nodata	&	\nodata	\\
6008.57	&	Fe I	&	3.88	&	$-$0.94	&	40	&	\nodata	&	121	&	\nodata	&	\nodata	&	\nodata	&	\nodata	&	\nodata	&	\nodata	&	\nodata	&	\nodata	\\
6015.25	&	Fe I	&	2.22	&	$-$4.66	&	\nodata	&	\nodata	&	26	&	\nodata	&	\nodata	&	\nodata	&	\nodata	&	\nodata	&	\nodata	&	\nodata	&	17	\\
6024.07	&	Fe I	&	4.55	&	$-$0.06	&	56	&	45	&	113	&	66	&	89	&	126	&	\nodata	&	110	&	97	&	\nodata	&	113	\\
6027.06	&	Fe I	&	4.07	&	$-$1.14	&	\nodata	&	\nodata	&	80	&	\nodata	&	\nodata	&	73	&	49	&	90	&	75	&	\nodata	&	75	\\
6035.35	&	Fe I	&	4.29	&	$-$2.56	&	\nodata	&	\nodata	&	\nodata	&	\nodata	&	\nodata	&	\nodata	&	\nodata	&	\nodata	&	\nodata	&	\nodata	&	\nodata	\\
\enddata

\end{deluxetable}

\clearpage
\pagestyle{empty}
\setlength{\voffset}{2cm}

\tablenum{2c}
\tablecolumns{15}
\tablewidth{0pt}

\begin{deluxetable}{ccccccccccccccc}
\rotate
\tablecaption{Linelist and Equivalent Widths}
\tablehead{
\colhead{$\lambda$}	&
\colhead{Element}	&
\colhead{E.P.}	&
\colhead{log gf}	&
\colhead{35201}	&
\colhead{36282}     &
\colhead{38198}     &
\colhead{38232}     &
\colhead{39026}     &
\colhead{39048}     &
\colhead{39129}     &
\colhead{39392}     &
\colhead{41033}     &
\colhead{42508}     &
\colhead{43061}     \\
\colhead{(\AA)}	&
\colhead{}	&
\colhead{(eV)}	&
\colhead{}	&
\colhead{}      &
\colhead{}      &
\colhead{}      &
\colhead{}      &
\colhead{}      &
\colhead{}      &
\colhead{}      &
\colhead{}      &
\colhead{}      &
\colhead{}      &
\colhead{}
}

\startdata

6003.01	&	Fe I	&	3.88	&	$-$1.07	&	\nodata	&	\nodata	&	\nodata	&	\nodata	&	73	&	\nodata	&	\nodata	&	\nodata	&	\nodata	&	\nodata	&	\nodata	\\
6007.32	&	Ni I	&	1.68	&	$-$3.35	&	\nodata	&	\nodata	&	\nodata	&	\nodata	&	\nodata	&	\nodata	&	\nodata	&	\nodata	&	96	&	\nodata	&	\nodata	\\
6007.97	&	Fe I	&	4.65	&	$-$0.70	&	\nodata	&	\nodata	&	\nodata	&	\nodata	&	\nodata	&	58	&	\nodata	&	\nodata	&	66	&	\nodata	&	\nodata	\\
6008.57	&	Fe I	&	3.88	&	$-$0.94	&	\nodata	&	\nodata	&	\nodata	&	\nodata	&	\nodata	&	\nodata	&	\nodata	&	\nodata	&	131	&	\nodata	&	\nodata	\\
6015.25	&	Fe I	&	2.22	&	$-$4.66	&	23	&	\nodata	&	\nodata	&	\nodata	&	11	&	36	&	\nodata	&	28	&	34	&	\nodata	&	\nodata	\\
6024.07	&	Fe I	&	4.55	&	$-$0.06	&	108	&	61	&	89	&	90	&	87	&	141	&	\nodata	&	130	&	134	&	91	&	\nodata	\\
6027.06	&	Fe I	&	4.07	&	$-$1.14	&	71	&	\nodata	&	51	&	74	&	53	&	111	&	\nodata	&	93	&	96	&	33	&	\nodata	\\
6035.35	&	Fe I	&	4.29	&	$-$2.56	&	\nodata	&	\nodata	&	\nodata	&	\nodata	&	\nodata	&	\nodata	&	\nodata	&	\nodata	&	\nodata	&	\nodata	&	19	\\
\enddata

\end{deluxetable}

\clearpage
\pagestyle{empty}
\setlength{\voffset}{2cm}

\tablenum{2d}
\tablecolumns{15}
\tablewidth{0pt}

\begin{deluxetable}{ccccccccccccccc}
\rotate
\tablecaption{Linelist and Equivalent Widths}
\tablehead{
\colhead{$\lambda$}	&
\colhead{Element}	&
\colhead{E.P.}	&
\colhead{log gf}	&
\colhead{43389}	&
\colhead{45358}     &
\colhead{45485}     &
\colhead{46121}     &
\colhead{47215}     &
\colhead{48116}     &
\colhead{48323}     &
\colhead{49013}     &
\colhead{49037}     &
\colhead{51021}     &
\colhead{51080}     \\
\colhead{(\AA)}	&
\colhead{}	&
\colhead{(eV)}	&
\colhead{}	&
\colhead{}      &
\colhead{}      &
\colhead{}      &
\colhead{}      &
\colhead{}      &
\colhead{}      &
\colhead{}      &
\colhead{}      &
\colhead{}      &
\colhead{}      &
\colhead{}
}

\startdata

6003.01	&	Fe I	&	3.88	&	$-$1.07	&	82	&	\nodata	&	\nodata	&	\nodata	&	\nodata	&	\nodata	&	\nodata	&	\nodata	&	\nodata	&	\nodata	&	\nodata	\\
6007.32	&	Ni I	&	1.68	&	$-$3.35	&	\nodata	&	\nodata	&	\nodata	&	91	&	\nodata	&	\nodata	&	\nodata	&	\nodata	&	\nodata	&	\nodata	&	\nodata	\\
6007.97	&	Fe I	&	4.65	&	$-$0.70	&	37	&	\nodata	&	\nodata	&	\nodata	&	\nodata	&	84	&	60	&	\nodata	&	\nodata	&	49	&	\nodata	\\
6008.57	&	Fe I	&	3.88	&	$-$0.94	&	89	&	\nodata	&	\nodata	&	\nodata	&	\nodata	&	\nodata	&	\nodata	&	\nodata	&	\nodata	&	90	&	\nodata	\\
6015.25	&	Fe I	&	2.22	&	$-$4.66	&	\nodata	&	\nodata	&	22	&	\nodata	&	32	&	\nodata	&	\nodata	&	8	&	\nodata	&	\nodata	&	\nodata	\\
6024.07	&	Fe I	&	4.55	&	$-$0.06	&	101	&	117	&	114	&	105	&	144	&	129	&	134	&	71	&	50	&	\nodata	&	102	\\
6027.06	&	Fe I	&	4.07	&	$-$1.14	&	64	&	74	&	85	&	90	&	\nodata	&	93	&	95	&	53	&	19	&	62	&	79	\\
6035.35	&	Fe I	&	4.29	&	$-$2.56	&	\nodata	&	\nodata	&	\nodata	&	\nodata	&	\nodata	&	\nodata	&	\nodata	&	\nodata	&	\nodata	&	\nodata	&	\nodata	\\
\enddata

\end{deluxetable}

\clearpage
\pagestyle{empty}
\setlength{\voffset}{2cm}

\tablenum{2e}
\tablecolumns{15}
\tablewidth{0pt}

\begin{deluxetable}{ccccccccccccccc}
\rotate
\tablecaption{Linelist and Equivalent Widths}
\tablehead{
\colhead{$\lambda$}	&
\colhead{Element}	&
\colhead{E.P.}	&
\colhead{log gf}	&
\colhead{51132}	&
\colhead{54022}     &
\colhead{54105}     &
\colhead{55101}     &
\colhead{55111}     &
\colhead{55142}     &
\colhead{56040}     &
\colhead{57010}     &
\colhead{59036}     &
\colhead{59090}     &
\colhead{59094}     \\
\colhead{(\AA)}	&
\colhead{}	&
\colhead{(eV)}	&
\colhead{}	&
\colhead{}      &
\colhead{}      &
\colhead{}      &
\colhead{}      &
\colhead{}      &
\colhead{}      &
\colhead{}      &
\colhead{}      &
\colhead{}      &
\colhead{}      &
\colhead{}
}

\startdata

6003.01	&	Fe I	&	3.88	&	$-$1.07	&	\nodata	&	\nodata	&	\nodata	&	101	&	\nodata	&	\nodata	&	\nodata	&	\nodata	&	\nodata	&	49	&	\nodata	\\
6007.32	&	Ni I	&	1.68	&	$-$3.35	&	\nodata	&	\nodata	&	\nodata	&	\nodata	&	\nodata	&	\nodata	&	\nodata	&	\nodata	&	\nodata	&	\nodata	&	\nodata	\\
6007.97	&	Fe I	&	4.65	&	$-$0.70	&	53	&	\nodata	&	\nodata	&	\nodata	&	\nodata	&	\nodata	&	\nodata	&	\nodata	&	\nodata	&	\nodata	&	\nodata	\\
6008.57	&	Fe I	&	3.88	&	$-$0.94	&	114	&	\nodata	&	\nodata	&	\nodata	&	\nodata	&	\nodata	&	\nodata	&	\nodata	&	\nodata	&	\nodata	&	\nodata	\\
6015.25	&	Fe I	&	2.22	&	$-$4.66	&	22	&	\nodata	&	\nodata	&	25	&	\nodata	&	\nodata	&	\nodata	&	7	&	\nodata	&	\nodata	&	\nodata	\\
6024.07	&	Fe I	&	4.55	&	$-$0.06	&	123	&	135	&	120	&	119	&	91	&	\nodata	&	56	&	82	&	74	&	65	&	39	\\
6027.06	&	Fe I	&	4.07	&	$-$1.14	&	85	&	112	&	84	&	88	&	54	&	\nodata	&	31	&	51	&	27	&	45	&	\nodata	\\
6035.35	&	Fe I	&	4.29	&	$-$2.56	&	\nodata	&	\nodata	&	\nodata	&	\nodata	&	\nodata	&	\nodata	&	\nodata	&	\nodata	&	\nodata	&	\nodata	&	\nodata	\\
\enddata

\end{deluxetable}

\clearpage
\pagestyle{empty}
\setlength{\voffset}{2cm}

\tablenum{2f}
\tablecolumns{15}
\tablewidth{0pt}

\begin{deluxetable}{ccccccccccccccc}
\rotate
\tablecaption{Linelist and Equivalent Widths}
\tablehead{
\colhead{$\lambda$}	&
\colhead{Element}	&
\colhead{E.P.}	&
\colhead{log gf}	&
\colhead{60066}	&
\colhead{63021}     &
\colhead{65046}     &
\colhead{66047}     &
\colhead{66054}     &
\colhead{69027}     &
\colhead{76038}     &
\colhead{77025}     &
\colhead{80029}     &
\colhead{81018}     &
\colhead{85027}     \\
\colhead{(\AA)}	&
\colhead{}	&
\colhead{(eV)}	&
\colhead{}	&
\colhead{}      &
\colhead{}      &
\colhead{}      &
\colhead{}      &
\colhead{}      &
\colhead{}      &
\colhead{}      &
\colhead{}      &
\colhead{}      &
\colhead{}      &
\colhead{}
}

\startdata

6003.01	&	Fe I	&	3.88	&	$-$1.07	&	\nodata	&	\nodata	&	\nodata	&	\nodata	&	\nodata	&	\nodata	&	\nodata	&	81	&	\nodata	&	\nodata	&	\nodata	\\
6007.32	&	Ni I	&	1.68	&	$-$3.35	&	\nodata	&	\nodata	&	\nodata	&	\nodata	&	\nodata	&	\nodata	&	\nodata	&	\nodata	&	\nodata	&	\nodata	&	\nodata	\\
6007.97	&	Fe I	&	4.65	&	$-$0.70	&	\nodata	&	\nodata	&	\nodata	&	\nodata	&	\nodata	&	\nodata	&	\nodata	&	\nodata	&	\nodata	&	\nodata	&	\nodata	\\
6008.57	&	Fe I	&	3.88	&	$-$0.94	&	\nodata	&	\nodata	&	\nodata	&	\nodata	&	\nodata	&	\nodata	&	\nodata	&	\nodata	&	\nodata	&	\nodata	&	\nodata	\\
6015.25	&	Fe I	&	2.22	&	$-$4.66	&	\nodata	&	\nodata	&	\nodata	&	\nodata	&	\nodata	&	\nodata	&	13	&	\nodata	&	11	&	\nodata	&	\nodata	\\
6024.07	&	Fe I	&	4.55	&	$-$0.06	&	74	&	71	&	39	&	89	&	\nodata	&	\nodata	&	106	&	72	&	76	&	\nodata	&	62	\\
6027.06	&	Fe I	&	4.07	&	$-$1.14	&	\nodata	&	61	&	\nodata	&	56	&	\nodata	&	\nodata	&	81	&	\nodata	&	43	&	42	&	45	\\
6035.35	&	Fe I	&	4.29	&	$-$2.56	&	\nodata	&	\nodata	&	\nodata	&	\nodata	&	\nodata	&	\nodata	&	\nodata	&	\nodata	&	\nodata	&	\nodata	&	\nodata	\\
\enddata

\end{deluxetable}

\clearpage
\pagestyle{empty}
\setlength{\voffset}{2cm}

\tablenum{3}
\tablecolumns{16}
\tablewidth{0pt}

\begin{deluxetable}{cccccccccccccccc}
\tabletypesize{\scriptsize}
\rotate
\tablecaption{Derived Abundances}
\tablehead{
\colhead{Star}      &
\colhead{[FeI/H]}      &
\colhead{[FeII/H]}      &
\colhead{[Fe/H]}      &
\colhead{[NaI/Fe]}      &
\colhead{[AlI/Fe]}      &
\colhead{[CaI/Fe]}      &
\colhead{[ScI/Fe]}      &
\colhead{[ScII/Fe]}      &
\colhead{[Sc/Fe]}      &
\colhead{[TiI/Fe]}      &
\colhead{[TiII/Fe]}      &
\colhead{[Ti/Fe]}      &
\colhead{[NiI/Fe]}      &
\colhead{[LaII/Fe]}      &
\colhead{[EuII/Fe]}     \\
\colhead{LEID}	&
\colhead{}      &
\colhead{}      &
\colhead{Avg.}      &
\colhead{}      &
\colhead{}      &
\colhead{}      &
\colhead{}      &
\colhead{}      &
\colhead{Avg.}      &
\colhead{}      &
\colhead{}      &
\colhead{Avg.}      &
\colhead{}      &
\colhead{}      &
\colhead{}       
}

\startdata

9	&	$-$1.47	&	$-$1.34	&	$-$1.41	&	$-$0.08	&	$+$0.06	&	$+$0.26	&	$-$0.07	&	$+$0.21	&	$+$0.07	&	$+$0.08	&	$-$0.03	&	$+$0.03	&	$-$0.06	&	$-$0.41	&	$-$0.41	\\
6017	&	$-$1.41	&	$-$1.03	&	$-$1.22	&	$+$0.36	&	$+$1.01	&	$+$0.20	&	\nodata	&	\nodata	&	\nodata	&	$-$0.07	&	$-$0.13	&	$-$0.10	&	$-$0.21	&	$+$0.89	&	$-$0.16	\\
12013	&	$-$1.37	&	$-$1.24	&	$-$1.31	&	$+$0.36	&	$+$0.24	&	$+$0.40	&	$-$0.10	&	\nodata	&	$-$0.10	&	$+$0.21	&	$+$0.20	&	$+$0.21	&	$-$0.03	&	$+$0.29	&	$-$0.46	\\
15023	&	$-$1.87	&	$-$1.73	&	$-$1.80	&	$-$0.50	&	$+$0.18	&	$+$0.20	&	\nodata	&	$+$0.12	&	$+$0.12	&	$+$0.04	&	$+$0.38	&	$+$0.21	&	$-$0.04	&	$-$0.03	&	$-$0.21	\\
16015	&	$-$1.83	&	$-$1.76	&	$-$1.80	&	$-$0.66	&	$+$0.13	&	$+$0.30	&	\nodata	&	\nodata	&	\nodata	&	$+$0.16	&	$+$0.30	&	$+$0.23	&	$-$0.10	&	$+$0.15	&	$+$0.18	\\
17015	&	$-$1.82	&	\nodata	&	$-$1.82	&	\nodata	&	\nodata	&	$+$0.38	&	\nodata	&	$+$0.39	&	$+$0.39	&	$+$0.09	&	$+$0.23	&	$+$0.16	&	$-$0.07	&	$-$0.07	&	$+$0.02	\\
19062	&	$-$1.70	&	$-$1.80	&	$-$1.75	&	$+$0.07	&	\nodata	&	$+$0.46	&	$+$0.27	&	\nodata	&	$+$0.27	&	$+$0.21	&	$+$0.45	&	$+$0.33	&	$+$0.04	&	$+$0.58	&	$+$0.27	\\
22037	&	$-$1.88	&	$-$1.63	&	$-$1.76	&	$-$0.09	&	$+$0.82	&	$+$0.23	&	\nodata	&	$+$0.11	&	$+$0.11	&	$+$0.07	&	$+$0.23	&	$+$0.15	&	$-$0.11	&	$-$0.40	&	$+$0.30	\\
\enddata

\end{deluxetable}

\clearpage

\tablenum{4}
\tablecolumns{5}
\tablewidth{0pt}

\begin{deluxetable}{lcccc}
\tablecaption{Abundance Sensitivity to Model Atmosphere Parameters}
\tablehead{
\colhead{Element}	&
\colhead{$\Delta$T$_{\rm eff}$$\pm$100}	&
\colhead{$\Delta$log g$\pm$0.30}	&
\colhead{$\Delta$[M/H]$\pm$0.30}	&
\colhead{$\Delta$V$_{\rm t}$$\pm$0.30}	\\
\colhead{}	&
\colhead{(K)}	&
\colhead{(cm s$^{\rm -2}$)}	&
\colhead{(dex)}	&
\colhead{(km s$^{\rm -1}$)}
}

\startdata

\hline
\multicolumn{5}{c}{[Fe/H]$\approx$--2.0}	\\
\hline
Fe I	&	$\pm$	0.14	&	$\mp$	0.01	&	$\mp$	0.04	&	$\mp$	0.06	\\
Fe II	&	$\mp$	0.02	&	$\pm$	0.11	&	$\pm$	0.06	&	$\mp$	0.01	\\
Na I	&	$\pm$	0.08	&	$\mp$	0.03	&	$\mp$	0.03	&	$\pm$	0.00	\\
Al I	&	$\pm$	0.07	&	$\mp$	0.02	&	$\mp$	0.03	&	$\mp$	0.02	\\
Ca I	&	$\pm$	0.12	&	$\mp$	0.04	&	$\mp$	0.06	&	$\mp$	0.10	\\
Ti I	&	$\pm$	0.19	&	$\mp$	0.03	&	$\mp$	0.05	&	$\mp$	0.01	\\
Ti II	&	$\mp$	0.04	&	$\pm$	0.12	&	$\pm$	0.08	&	$\mp$	0.03	\\
\enddata

\end{deluxetable}

\clearpage

\tablenum{5}
\tablecolumns{2}
\tablewidth{0pt}

\begin{deluxetable}{ll}
\tablecaption{Literature References for Figures \ref{f11}--\ref{f16}}
\tablehead{
\colhead{Object}	&
\colhead{Reference}
}

\startdata

Thin/Thick Disk	&	Bensby et al. (2003)	\\
Thin/Thick Disk &       Bensby et al. (2005)    \\
Thin/Thick Disk &       Fulbright et al. (2000)    \\
Thin/Thick Disk &       Fulbright et al. (2007)	\\
Thin/Thick Disk &       Reddy et al. (2003)	\\
Thin/Thick Disk	&	Simmerer et al. (2004)	\\
Halo	&	Fulbright et al. (2000)	\\
Halo	&	Reddy et al. (2006)	\\
Halo	&	Simmerer et al. (2004)	\\
Bulge	&	McWilliam \& Rich (1994)	\\
Bulge	&	Fulbright et al. (2007)	\\
Bulge	&	Lecureur et al. (2007)	\\
Dwarf Spheroidals	&	Shetrone et al. (2001)	\\
Dwarf Spheroidals	&	Shetrone et al. (2003)	\\
Sagittarius Dwarf	&	Sbordone et al. (2007)	\\
Globular Cluster (M4)   &       Ivans et al. (1999)     \\
Globular Cluster (M5)   &       Ivans et al. (2001)     \\
Globular Cluster (M13)  &       Sneden et al. (2004)   \\
Globular Cluster (M13)  &       Johnson et al. (2005)    \\
Globular Cluster (M15\tablenotemark{a})  &       Sneden et al. (1997)	\\
Globular Cluster (M68\tablenotemark{a})  &       Lee et al. (2005)       \\
Globular Cluster (M71)  &       Ram{\'{\i}}rez \& Cohen (2002)   \\
Globular Cluster (M80)  &       Cavallo et al. (2004)   \\
Globular Cluster (47 Tuc)	&	Carretta et al. (2004)	\\
Globular Cluster (47 Tuc)	&	James et al. (2004)	\\
Globular Cluster (NGC 288)	&	Shetrone \& Keane (2000)	\\
Globular Cluster (NGC 362)	&	Shetrone \& Keane (2000)	\\

\enddata

\tablenotetext{a}{These two clusters have had their measured [Fe/H] values 
shifted by +0.2 dex to fit in Figures \ref{f11}--\ref{f14}.  These shifts 
were not applied to the displayed [X/Fe] abundances.}

\end{deluxetable}


\end{document}